\documentclass[sigconf]{acmart}
\usepackage{amsmath,amsfonts}
\usepackage{algorithmic}
\usepackage{graphicx}
\usepackage{textcomp}
\usepackage{tabularx}
\newcolumntype{Y}{>{\centering\arraybackslash}X}
\newcolumntype{R}{>{\raggedleft\arraybackslash}X}

\usepackage{caption}
\usepackage{subcaption}
\usepackage{multirow}
\usepackage{multicol}
\usepackage{makecell}
\usepackage{xcolor} 
\usepackage{xurl}
\usepackage{todonotes}
\usepackage{booktabs}
\usepackage{censor}
\usepackage{stfloats}
\AtBeginDocument{%
  \providecommand\BibTeX{{%
    \normalfont B\kern-0.5em{\scshape i\kern-0.25em b}\kern-0.8em\TeX}}}

\setcopyright{rightsretained}
\copyrightyear{2024}
\acmYear{2024}
\setcopyright{acmlicensed}\acmConference[NSPW '24]{New Security Paradigms Workshop}{September 16--19, 2024}{Bedford, PA, USA}
\acmBooktitle{New Security Paradigms Workshop (NSPW '24), September 16--19, 2024, Bedford, PA, USA}
\acmDOI{10.1145/3703465.3703469}
\acmISBN{979-8-4007-1128-2/24/09}

\begin{document}

\title{`Debunk-It-Yourself': Health Professionals' Strategies for Responding to Misinformation on TikTok}
% \todo{SD: Title Suggestion: Debunk-It-Yourself: Health Professionals' Strategies for Combating Misinformation on TikTok?. We can probably use a lighter word than Combat?}

\author{Filipo Sharevski}
\affiliation{%
  \institution{DePaul University}
  \streetaddress{243 S Wabash Ave}
  \city{Chicago, IL}
  \country{United States}}
\email{fsharevs@depaul.edu}

\author{Jennifer Vander Loop}
\affiliation{%
  \institution{DePaul University}
  \streetaddress{243 S Wabash Ave}
  \city{Chicago, IL}
  \country{United States}}
\email{jvande27@depaul.edu}

\author{Amy Devine}
\affiliation{%
  \institution{DePaul University}
  \streetaddress{243 S Wabash Ave}
  \city{Chicago, IL}
  \country{United States}}
\email{adevine@depaul.edu}

\author{Peter Jachim}
\affiliation{%
  \institution{DePaul University}
  \streetaddress{243 S Wabash Ave}
  \city{Chicago, IL}
  \country{United States}}
\email{pjachim@depaul.edu}

\author{Sanchari Das}
\affiliation{%
  \institution{George Mason University}
  \streetaddress{}
  \city{Fairfax, VA}
  \country{United States}}
\email{sdas35@gmu.edu}

\renewcommand{\shortauthors}{F. Sharevski, J. Vander Loop, A. Devine, P.Jachim, S. Das}

\begin{abstract}
Misinformation is ``sticky'' in nature, requiring a considerable effort to undo its influence. One such effort is debunking or exposing the falsity of information. As an abundance of misinformation is on social media, platforms do bear some debunking responsibility in order to preserve their trustworthiness as information providers. A subject of interpretation, platforms poorly meet this responsibility and allow dangerous health misinformation to influence many of their users. This open route to harm did not sit well with health professional users, who recently decided to take the debunking into their own hands. To study this individual debunking effort --- which we call `\textit{Debunk-It-Yourself}' --- we conducted an exploratory survey \textit{n}=14 health professionals who wage a misinformation counter-influence campaign through videos on TikTok. We focused on two topics, nutrition and mental health, which are the ones most often subjected to misinformation on the platform. Our   analysis reveals that the counterinfluence follows a common process of initiation, selection, creation, and ``stitching'' or duetting a debunking video with a misinformation video. The `\textit{Debunk-It-Yourself}' effort was underpinned by three unique aspects: (i) it targets trending misinformation claims perceived to be of direct harm to people's health; (ii) it offers a symmetric response to the misinformation; and (iii) it is strictly based on scientific evidence and claimed clinical experience. Contrasting the `\textit{Debunk-It-Yourself}' effort with the one TikTok and other platforms (reluctantly) put in moderation, we offer recommendations for a structured response against the misinformation's influence by the users themselves.

% Using affordances for direct response to content such as ``duets'' (stitching) and tagging on TikTok, these users started undoing the misinformation's influence by themselves. 

% \todo{ recommendations?}

\end{abstract}





\maketitle

\section{Introduction}

In the digital age, the rapid dissemination of false information (or \textit{misinformation}), often with an intent to mislead (or \textit{disinformation}), has necessitated the development of strategies to counteract the spread of incorrect claims. Debunking, defined as the act of exposing the incorrectness of these claims, has become an essential tool in protecting public discourse and maintaining trust in both digital institutions such as social media platforms and actual institutions such as health authorities ~\cite{Lewandowsky2020}. Debunking as a response to misinformation is not merely about correcting false information but involves a complex interaction of various elements within the online information ecosystem, particularly social media platforms.

Despite the availability of resources like the ``Debunking Handbook'' which provides guidelines for effective misinformation correction ~\cite{Lewandowsky2020}, the actual application of debunking in online environments has encountered significant challenges~\cite{Kozyreva2023}. Social media platforms, primarily  due to network effects, have been central to the spread of misinformation ~\cite{anderson2014privacy}. Their business models, which prioritize engagement, content virality, and user lock-in, often conflict with the imperative of maintaining accurate and reliable information. As a result, these platforms have largely opted for moderation techniques that involve less direct confrontation of false claims, such as the use of warning labels linked to third-party information~\cite{youtube-misinfo, meta-misinfo, twitter-misinfo, tiktok-safety}. This indirect approach to handling misinformation through moderation rather than outright debunking has its repercussions. It allows platforms to avoid taking a definitive stance on the truthfulness of content, thereby sidestepping potential accusations of bias or censorship~\cite{Walter2020}. However, this also means that the responsibility of determining the veracity of information is shifted to external fact-checking services, raising questions about the effectiveness and transparency of the debunking process~\cite{Stewart2021, gettr-paper}.

Moreover, the evolution of social media platforms has introduced new dynamics into how misinformation is handled. The rise of TikTok, for example, with its algorithm-driven content distribution, has further complicated the landscape. The platform's focus on passive content consumption rather than active user interaction changes how misinformation is encountered and the visibility of any corrective measures like labels or community notes~\cite{Zeng2022, sharevski2023abortion}. The algorithmic trap of passive consumption necessitates a considerable supply of content so TikTok incentivizes and promotes creators that continuously post on the platform without much research whether the content is truthful and trustful. Borrowing the information economics explanation of why security is hard, TikTok knows that ignoring debunking makes the life easier for these creators, which in turn will help the platform lock-in users rather than just protecting them from falsehoods ~\cite{anderson2014privacy}. Amidst these challenges, a new form of response has emerged, particularly within the health misinformation domain. As active users on platforms like TikTok, health professionals have started directly addressing false claims through their content. This `\textit{Debunk-It-Yourself}' approach leverages their expertise and credibility to confront misinformation head-on, bypassing the platform's standard moderation practices~\cite{Raphael2022}. Such initiatives reflect a shift towards more proactive and personal involvement in debunking, tailored to the unique environment of modern social media platforms.

In particular, we were interested in mapping the entire `\textit{Debunk-It-Yourself}' process --- from how health professionals initiate and select candidate claims, to how they craft their debunking response, what experience they have with creators whose claims they debunked and with the algorithmic curation of content on TikTok, as well as the platform's coarse approach to moderation. We focused on topics related to nutrition and mental health because they are the main topics that ``influencers'' particularly exploit with unverified and dangerous content that includes self-diagnosis, treatment, weight loss advice, self-harm, and at-home recovery programs \cite{Harpal2023, PlushCare2022, cch2023}. As this is a new debunking paradigm, we didn't evaluate the effect of `\textit{Debunk-It-Yourself}' effort with other TikTok users, i.e. whether the debunking actually is effective in dislodging and false information that otherwise these users might have assumed as truthful and trustworthy. 

We instead opted to study the phenomenology of the individual debunking approach first in order to obtain a rich and comprehensive understanding of its inner workings in order to better inform a research agenda centered around the overall effects of an individual-first style of debunking. In other words, we saw methodologically sound to first describe what the `\textit{Debunk-It-Yourself}' effort \textit{is} about, and later to find out \textit{if} this effort works as the individuals doing the debunking might assume it would. This, in our view, includes future user evaluation studies that are not just focused on nutrition and mental health, but also other topics subject of misinformation and disinformation such as politics or public health; comparing and contrasting the `\textit{Debunk-It-Yourself}' effort with platform-provided or third-party debunking warning labels; or cross platform evaluation with users from TikTok, Instagram, or YouTube, to name a few. In this context, we see as equally important to specifically study the ``supply side'' of the `\textit{Debunk-It-Yourself}' effort as to learn the incentives, risks, interactions, side-effects, and motivation that drive individual users to debunk misinformation and disinformation on a social media platform  (or cease doing so, if they were already actively doing debunking before).

To identify the health professionals responding to nutrition and mental health misinformation, we used the TikTok API to collect datasets and find the most prominent and active ones, resulting in a population of $135$ accounts. After we vetted that they are accredited and/or certified health professionals, we contacted each of these individuals to participate in our study, and \textit{n}=14 agreed to participate in an anonymous exploratory survey. We chose the survey as an option instead of an interview to ensure that our results would not cause any further harm to these users and hamper their future debunking efforts on TikTok (e.g., avoid being targeted or directly harassed by those who got their content debunked). The answers to the exploratory survey (shown in the Appendix \ref{app:study-questions} were used to answer the following research question: \textit{How do prominent health professionals initiate, select, and debunk misinformation on TikTok relative to nutrition and mental health?}
 
\noindent \textbf{Findings}: Our analysis revealed a common debunking process that all of our participants followed, including steps for: (i) initiation; (ii) selection of a target video; (iii) creation of a debunking video; (iv) responding to the target video; and (v) post-debunking response management. The debunking process was usually initiated by a video containing false, misleading, or unvetted medical claims TikTok recommended to our participants' `For You Page' or their followers, tagging them to respond to trending falsehood. Unlike platform moderation, where the initiation is usually aided by automatic means, our participants were not actively employing any such measures, nor did they actively seek misinformation videos. Instead, they decided to debunk a video based either the perceived danger of the claims involved or if other users requested their help about a video they saw on TikTok.

\noindent \textbf{Scope and contributions:} While debunking misinformation is a scientifically well-mapped effort, it nonetheless falls short of materializing on social media platforms. The scope of our work was precisely to study this gap, filled by individual users in the absence of decisive action by platforms. Focusing on the most trending health misinformation topics on TikTok -- nutrition and mental health -- we uncover how health professionals, motivated to prevent individual harm, made a uniquely symmetric (e.g., content-to-content, instead of using just warnings or adding comments to content), crowd-based debunking effort so far. Our contributions, thus, are:

    \begin{itemize}
        \item A novel paradigm for debunking misinformation `\textit{Debunk-It-Yourself}' on TikTok where health professionals -- independent of the platform -- create content in response to content that contains false, misleading, or unvetted medical claims;
        
        \item A formal mapping of the `\textit{Debunk-It-Yourself}' process that includes distinct steps to initiate, select, create a video, respond, manage, and continuously sustain the debunking effort -- a primer exemplified on TikTok but of equal flexibility of implementation on other short video-based platforms such as Instagram or YouTube; 
        
        \item Recommendations for potential scenarios of utilizing the `\textit{Debunk-It-Yourself}' towards a coordinated, comprehensive, and platform-independent response against harmful health misinformation across the social media landscape.
    
    \end{itemize}

\section{Debunking -- Background}
The act of debunking, borne out by a rapid influx of false claims in recent times, carries the connotation of exposing the falseness of claims~\cite{Nichols2014}. The volumes of misinformation proliferated online set debunking apart as the pivotal \textit{response} to the threat of substantial harm to individuals and society caused by drowning the truth with what it ``feels like truth.'' However, demonstrating the wrongness of a thing or concept online proved not to be an easy affair. Despite having a scientifically developed and demonstratively easy-to-use ``debunking handbook''~\cite{Lewandowsky2020}, online places -- notably social media platforms -- balked when it came to truth arbitration. 

According to the ``debunking handbook''~\cite{Lewandowsky2020}, responding to misinformation is successfully done if one applies the following components: (1) stating the truth; (2) points to the misinformation;' (3) explain why misinformation is wrong; and (4) state the truth again.  When stating the truth, the handbook recommends doing it in a few clear words, not relying on a simple retraction (e.g., ``this claim is not true''), and providing a factual alternative that is an alternative that fills a causal ``gap'' in explaining what happened if the misinformation is corrected. The handbook recommends repeating the misinformation, only once, directly prior to the correction to avoid a situation where the repetition might make the misinformation appear true. Then, one should juxtapose the correction with the misinformation and provide details as to why the said claims are false. The explanation why the misinformation is wrong is best done if not only a factual alternative is provided but also by pointing out logical or argumentative fallacies underlying the misinformation. Finally, the handbook recommends the debunker to restate the fact again, so the fact is the last thing people process. 

The task of maintaining a reasonably ascertainable picture of reality proved to be at odds with the platforms' incentives to promote content that drives engagement while maintaining the image of impartial providers simply upholding users' freedom of expression. Debunking -- involving steps to point out the falsity of a claim and providing the actual truth -- was seen as too intrusive, and instead, moderation was chosen as a course of action~\cite{Kozyreva2023}. With moderation, platforms need not explicitly expose the falseness of claims but only ``offer'' warning labels linked to information from third parties that bears credibility relative to well-established historical, scientific, and health topics~\cite{youtube-misinfo, meta-misinfo, twitter-misinfo, tiktok-safety}. Platforms also carefully position themselves away from the role of ``debunker'' by placing the burden on third parties instead -- usually fact-checking services -- in case there are objections about bias, partiality, or simply mistaken attribution of falseness~\cite{Walter2020}. Considering the regulatory inaction to rein in on social media falsehoods, the provision of the actual truth fell out of their responsibility, bar the application of the warning labels to demonstrate a bare minimum social responsibility.   

Social media platforms choice to move away from debunking, nonetheless, did not mitigate accusations of favoritism, free speech infringement, and outright censorship~\cite{Stewart2021, gettr-paper}. As the labeling (and debunking) of contested claims involves value judgments, some platforms shifted the ``debunker'' role from third parties to the community of own non-expert users representing competing viewpoints to signal fairness through agreement or consensus when pointing out the falsity of claims and providing the actual truth~\cite{Drolsbach2023}. Harnessing the ``wisdom of the crowds'' and applying community notes instead of labels, though promising, have the disadvantage of disputes among users, which, evidence shows, find it hard to leave their preexisting political, ideological, or other biases aside when debunking claims on social media~\cite{Allen2021}. Transparency is another issue as the community notes do not offer much information about the whole ``debunking'' process~\cite{Jiang2023}.

Both the labels and community notes were designed to fit mostly an interaction-driven participatory model on social media where users mainly interact -- i.e., post, comment, share, and access content (claims) in a context defined by the topics or users of their interest~\cite{Bhandari2022}. The arrival of TikTok introduced a new, algorithmically-driven participatory model on social media where users are ``viewing'' -- instead of interacting with -- content that is divorced from context as it is opaquely curated by a central recommendation algorithm~\cite{Zeng2022}. The \textit{content} moderation thus shifted to \textit{visibility} moderation, and as such, rendered the labels and notes less applicable or relevant -- the attention of users was on the next short video in their feed instead of a label underneath the one currently presented to them~\cite{sharevski2023abortion}. TikTok does apply labels to signal minimum social responsibility, but they are either incongruently assigned based on hashtags or \textit{en masse} based on keywords, rendering them practically useless for any debunking purpose~\cite{Ling-Gummadi}.

\begin{table*}[bp]
\renewcommand{\arraystretch}{1.2}
\small
\caption{\textbf{Response to Misinformation: Evolution}}
\label{tab:response}
\centering
\begin{tabularx}{\linewidth}{|r|X|X|X|}
\hline
% & \multicolumn{2}{*}{\textbf{Platfrom Driven} &  \textbf{User Driven} \\\hline
\textbf{Attribute} & \textbf{Warning Labels} & \textbf{Community Notes} & \textbf{Debunk-it-Yourself} \\\hline
%%%%%%%%%%%%%%%%%%%%%%
% structure attributes
% %%%%%%%%%%%%%%%%%%%%%
Format & Brief warning with a link to a third-party website substantiated \textit{under} potentially misinformation content & Extended warning text with an optional link to a third-party website substantiated \textit{under} potentially misinformation content & Actual content with full warning, context, hashtags, links to scientific evidence substantiated \textit{next} to potentially misinformation content \\\hline
Function & Inform about well-established facts & Provide informative context & Debunk falsehoods \\\hline
Topics & Public health, politics, civic order & All (consensus-dependent) & Individual and public health \\\hline
Policy & Platform defined & Platform and user consensus & User defined \\\hline
Initiation & Platform moderators and fact-checkers & Users and non-expert fact-checkers & Individual expert users \\\hline
Selection & Algorithmically-aided \cite{Paudel2023, Ling-Gummadi} & Consensus-dependent \cite{Martel2023} & Content-dependent [Section \ref{sec:results}] \\\hline

%%%%%%%%%%%%%%%%%%%%
%response attributes
%%%%%%%%%%%%%%%%%%%%
Magnitude & \textit{Asymmetric} \cite{Gillespie2018} & \textit{Asymmetric} \cite{Allen2021} & \textit{Symmetric} [Section \ref{sec:results}] \\\hline 
% Algorithmic support & Detection and content promotion/demotion & Detection and content promotion/demotion & None \\\hline

%%%%%%%%%%%%%%%%%%%%%
%User attributes
%%%%%%%%%%%%%%%%%%%%%%
Debunker Profile & Platform & Non-expert fact checkers with diverse viewpoints, political or ideological \cite{Stewart2021} & Credentialed health professionals \cite{Raphael2022} \\\hline
Identity & Anonymous (Platform) & Anonymous (Crowd) & Non-Anonymous (User) \\\hline
Engagement & None \cite{twitter-misinfo} & Rating (helpfulness) \cite{twitter-community-notes} & Full \\\hline
Virality & High \cite{Papakyriakopoulos} & Medium \cite{Drolsbach2023, Saeed2022} & Low (for now) [Section \ref{sec:results}] \\\hline
% Reception & Ignoring, disagreement \cite{cose2022, Kirchner, Swire-Thompson} & Partial agreement \cite{Drolsbach2023} & User agreement, debunked content creator disagreement [Section \ref{sec:results}] \\\hline

 \end{tabularx}
\end{table*}

The algorithmically driven participation also has the propensity to drown the truth to what it feels like truth without much-needed external intervention, for example, social bots used in the past for misinformation amplification~\cite{woolley2018computational}. False and misleading claims packaged in short quickly found their way on TikTok~\cite{newsguard}, incentivized by the opportunity for quickly becoming viral, attracting followers, and eventually turning their creators into ``influencers'' on various topics~\cite{herrman2019tiktok}. Most notably, these topics revolve around health-related issues, and TikTok users could hardly avoid seeing videos about self-diagnosis, alternative treatments, nutrition, or anti-vaccination sentiment ~\cite{Southwick2021, Basch2021, Southerton2022, sharevski2023abortion}.

In the circumstances of lax moderation and unexposed false claims pertaining to health, health professionals decided to take the role of ``debunkers'' and, as users, create short videos that pointed out the falsity and offered the actual, scientific truth~\cite{Raphael2022}. This debunking response is \textit{new} and differs from everything seen so far on social media -- these health professional users are not anonymous like the users behind the community notes; they don't rely on third parties for credible information; they are experts in their fields and not professional fact-checkers; they don't abide by any platform policy about what claims are subject of debunking and in what way. More so, they do the debunking in actual videos instead of using platform affordances such as warning labels or community notes. In the absence of resolute platform moderation, these health professionals -- motivated to prevent direct and substantial harm to individuals, public health, and society -- have given rise to a platform-independent way of responding to misinformation.

We call this effort `\textit{Debunk-It-Yourself}' -- as a new approach where expertise, credibility, and clinical experience are channeled towards symmetric-in-content refutations to false claims. In the past, health professionals' responses were mainly asymmetric, delivered either through reply, comment, or sharing a social media post with contested claim~\cite{Bautista2021}. With features like ``stitching'' or \textit{duets} -- where content creators could push their content together with the one containing the claim~\cite{Southerton2022} -- the debunkers have the opportunity to respond in the same magnitude, pointing out the falsity of a claim, and offering the actual facts about a topic. These features also allow debunkers to apply and even experiment with the recommendations from the prescribed strategies for debunking~\cite{Lewandowsky2020} to determine the most effective way of correcting misinformation on TikTok~\cite{Prike2023}. While there has been extensive research on other debunking approaches, so far, the `\textit{Debunk-It-Yourself}' approach has not yet received much academic attention. To make a case for bringing attention to our study, we created a comparison between the three main types of responses to misinformation in Table \ref{tab:response}.

\section{Background Work}
\subsection{Platforms' Response to Misinformation}
Misinformation has a long history of exploiting the human belief processes and psychological shortcomings~\cite{Cortada, posetti2018short}. Until recently, responding to what \textit{feels like truth} was not of concern as offering the actual \textit{truth} seemed the intuitive way of correcting one's susceptibility to manipulation. But the technological advantages of social media -- notably the virality and the minimally regulated participation -- brought an unprecedented threat of drowning the truth with an unencumbered amount of misinformation~\cite{McIntyre-Post, feld2020telegraph}. Offering the truth, it quickly became apparent, would not prevent the widespread manipulation of public belief, as social media participation is structured to allow for perpetuating, asymmetric exposure to content that feels like truth (i.e. content that on a first inspection appears correct but after a closer inspection one could reveal logical fallacies and incorrect claims)~\cite{cose2022}. Misinformation is also ``sticky'' in nature or continues to influence one's thinking even after they receive and accept the truth as correction~\cite{johnson1994sources}. 

Napoli et al. mention that social media platforms, in these circumstances, couldn't maintain their self-conception of impartial, non-corrective technology providers for long~\cite{Napoli2017} as misinformation started to shape decisions with harmful consequences relative to both individual and public health as well as civic order and interpersonal relations. Platforms had to do something about correcting misinformation, possessing the means, but they lacked (and still do) the necessary motivation to do so~\cite{anderson2019truth}. Correcting misinformation naturally entails \textit{truth arbitration}, a role that is at odds with their economic incentives (moderating content that drives engagement), reputational incentives (avoiding accusations of partisan censorship), and ideological incentives (upholding freedom of expression). But social media platforms, as key institutions in our public sphere, nonetheless knew they must preserve their trustworthiness and make some strides relative to how misinformation reaches wider audiences, if not how it exploits their psychological flaws.  

The tension between incentives and the responsibility to promote the truth forced the platforms to steer away from overt corrections and instead respond with a rather opaque course of \textit{moderation}. The moderation of misinformation included removing outright falsehoods (i.e., hard moderation), leaving misinformation to stay on the platform, and adding informative labels that hinted that such content might feel like truth (i.e., soft moderation). YouTube developed and implemented \textit{information panels}, Twitter (now X) used \textit{labels}, Meta provided \textit{warning labels}, and TikTok followed suit with \textit{content labels}, all intended to discourage users from choosing what \textit{feels like truth} and instead \textit{get the actual truth} about a topic. The goal of these labels, platforms argued, was to inform and contextualize content by sharing timely information or credible content from third-party sources for well-established historical, scientific, and health topics that are often subject to misinformation~\cite{youtube-misinfo, meta-misinfo, twitter-misinfo, tiktok-safety}.

Despite their minimally intrusive design~\cite{context2022}, the labels had an unwelcome spotlight among social media users. Users mainly saw a politically ulterior motive for implementing the labels as an attempt of the platforms to exercise their \textit{truth arbitration} option in favor of a perceived left-leaning ideology and silence right-leaning or other unpopular opinions (US political dichotomy, as most of the mainstream platforms mentioned above are US companies)~\cite{Sharevski-Rumors}. The warning labels were not only seen as intrusive to the freedom of speech~\cite{gettr-paper}, but they became -- for some time -- markers of \textit{actual truth} of any content they substantiated, instead of the other way around~\cite{Swire-Thompson}. Nonetheless, platforms maintained the labels as they still are advantageous for the platforms' ultimate goal of minimal intervention. Labels blend with the user interface aesthetics, contain vague text, place the \textit{truth arbitration} burden on a third party such as a fact-checking service or a health authority, and most importantly, they allow the misinformation content to remain available (and be algorithmically promoted) on the platform~\cite{Sharevski-nspw2023}. 

A disadvantage for the warning labels, however, is that they were formatted or placed in the user interface and remained directly associated with the platform as the one that ultimately decides what content is labeled as potential misinformation~\cite{Gillespie2018}. Platforms, notably Twitter, realized that removing the image of ``overtly imposing,'' ``biased,'' and ``punitive'' \textit{truth arbitrator} necessitates shifting the association from the platform directly to the users. The warning labels then evolved to \textit{community notes} where users -- instead of the platform moderators -- were to add context and correct inaccurate or misleading content. The shift to crowdsourcing ``fact checking'' by non-expert users was seen as the way forward in responding to misinformation, especially because it was designed to involve users with both left-leaning and right-leaning political worldviews in the process of target content selection, crafting the text of the response to a misinformative content, and assigning the community notes themselves~\cite{Saeed2022}. The whole concept is consensus-dependent so the community must first reach an agreement for each misinformation topic -- which they choose based on a broader policy targeting falsehoods instead of a selective labeling policy targeting only public health, elections, and public affairs -- to be addressed and contextualized to the other users~\cite{twitter-community-notes}. 

The community notes, so far, have avoided being politicized as the labels were, and users seem to receive them in a better light, acknowledging the effort for harnessing the ``wisdom of the crowds''~\cite{Drolsbach2023}. The community notes, unlike the labels, moved the response further from simply providing context and credible information -- the \textit{actual truth} against what \textit{felt like truth} -- to instances of \textit{debunking} or direct refutation of falsehoods~\cite{Bernecker}. This is different from just offering users two narratives around an issue and leaving them to decide what to pick to believe in -- the debunking entails two steps where first, it is explained why one narrative is false or misleading, and second, what is the actual truth instead~\cite{Lewandowsky2020}. Because the actual truth could be more complicated, i.e., feeling less appealing than sensational misinformation, the debunking done through the community notes requires additional investments in complex ideas in easily imagined and readily comprehensible terms for the wider audiences, resulting in extended warning texts and longer ``labels'' that substantiate contested posts.

Developed after the pronounced period of misinformation tumult on social media where platforms were caught unprepared for the influx of falsehoods (US elections, global pandemic), however, the community notes haven't really been put to the test and time will tell whether the consensus-driven response will resist the temptations of contested narratives that feel like truth. The length and variability of the community notes might also be subject to habituation (i.e., attenuation of a user response with multiple exposures), leading to situations where users simply ignore the context provided and the debunked misinformation. Further, The agreements are not devoid of biases, and research shows that non-expert users bring their partisan biases in crowdsourced fact-checking on social media~\cite{Allen2021}. Another equally concerning thing is that the community notes, like the labels, are anonymous and assigned by non-experts -- ordinary users do not know who these users are, what qualifications they have, and how the credibility of truthful sources is ascertained for both well-established and emerging historical, scientific, and health topics that are often subject to, or have the potential to be subjected, to misinformation. 

\subsection{Individual Users' Response to Misinformation}
Both the labels and the community notes are platform interventions and, as such, are shaped by the platform policies defining misinformation and where the platform stands relative to their self-conceptions as simply content providers. Twitter and Meta, for example, reverted and changed their misinformation policies several times in the past years, affecting what gets contextualized, labeled, removed, or algorithmically promoted/demoted~\cite{Roth2022-Twitter, Davies2023-Meta}. The instability of the \textit{truth arbitration} or moderation -- whether sharing actual truth or debunking falsehoods -- sends a clear signal that platforms are not overly concerned with the unrelenting drowning of truth with misinformation. Here, Altay et al. argue that misinformation (i) is not a social media problem but a legacy media problem; (ii) does not come in large quantities; and (iii) it doesn't harm people much as it doesn't much change their behavior~\cite{Altay2023}. 

While this argument merits an academic discussion, working nicely along the lines of the tobacco disinformation playbook to exculpate platforms from implication~\cite{michaels2020triumph}, individual users felt they need to do debunking themselves -- outside of platform-controlled crowdsourcing -- if the platforms won't do it. This is nothing new, as users often debunk through comments, replies, and discussions on social media platforms to dissuade others from accepting content that feels like truth. Debunking through engagement exists and exists together with the labels and the community notes, and often does link third-party credible information, though chosen by the users and not the platform moderators~\cite{Jiang2018}. But being directly corrected by another user in the comments section of the platform, evidence shows, decreases the quality and increases the partisan slant and language toxicity of the users' subsequent content posts~\cite{Mosleh2021}.

Individual users concerned with misinformation recently shifted towards creating content, instead of commenting, that debunks falsehoods found on social media~\cite{Bhargava2023}. This shift was facilitated by the shift in the platform affordances, particularly the proliferation of short videos and algorithmically tailored content consumption, characteristic for TikTok, Instagram Reels, or YouTube shorts~\cite{Bhandari2022}. The ``viewing'' -- rather than the traditional ``interacting'' experience on these platforms allows for users to ``compete'' with others on creatively crafting content that, based on users' personalized consumption~\cite{Klug2021}, is pushed to the trending section and recommended to others on the platform~\cite{Boeker2022}. Adapting to both the novel social media affordances and audiences, many took the advantage to push misinformation videos on these platforms~\cite{Hsu, newsguard}.  

The content that feels like truth is mainly disseminated around health-related topics or conspiracy theories~\cite{Basch2021, sharevski2023abortion, Kang, Grandinetti2022}. True to the minimal involvement in moderating falsehoods and responding to misinformation, platforms -- TikTok in particular -- don't do much to remove this content or label it correctly. Evidence shows that TikTok moderates videos based on hashtags included in the description without an in-depth analysis of the content~\cite{Ling-Gummadi} or bias against non-mainstream commentary in videos~\cite{Zeng2022}. With this new way of participation, the threat of drowning the truth expanded to include algorithmically recommended short videos, in addition to -- so far, dominant -- memes and textual content. Users could still comment on videos in response to the falsehoods or misleading claims stated in them, but the comments as the perception of engagement drowns correction signals -- the number of views, followers, and likes as signals of the videos' credibility~\cite{Cheng2024}. 

While debunking conspiratorial narratives is a consuming and convoluted effort~\cite{Dentith2021}, exposing the falseness of health-related claims could be relatively easily done within a short video and posted on social media~\cite{Bhargava2023}. In the absence of adequate platform moderation, users realized they could create such videos and post them as a direct refutation of what they saw as misinformation. Engagement was key, so users realized that the debunking content they produced faced an uphill battle against the views, followers, and likes -- usually on a virality-level due to their sensationalist nature preferred by the algorithmic recommendation~\cite{Chu2022} -- of the misinformation claims~\cite{Chao_Wang_Yu_2021}. While seemingly discouraging for most users, these circumstances singled out healthcare professionals as the ones who were ready to fight the battle against misinformation through short videos, most notably on TikTok~\cite{Raphael2022}.    

Healthcare professionals, including physicians and nurses, are motivated to debunk misinformation because they want to stand up for what is right or because they want to keep people safe~\cite{Bautista2021, Oktavianus2023}. Previous studies have shown that healthcare professionals exposed misinformation through commenting, replying, and posting on Twitter or Facebook~\cite{Bautista2021b}. Recently, they moved their debunking on TikTok, harnessing the unique platform affordance of ``stitching'' a video or creating a \textit{duet} -- a response video that allows users to reply directly to a video post with a video of their own~\cite{Southerton2022}. Aware that responding to misinformation is a controversial issue and suffers from polarization, bias, and refutation relativization, health professionals took a professional approach, projecting an image of expertise to be perceived as credible by other users ~\cite{Pretorius2022}. They also made sure to shape their response to misinformation in line with guidance from their professional organizations \cite{NurseGuidelines, amaGuidelines}. However useful, this nonetheless falls short of explicitly discovering and selecting misinformation to debunk by this group.

Healthcare professionals on social media are expected to adhere to the standards of their medical profession and provide truthful information when utilizing their credentials to present factual information~\cite{Schluger2022}. Upholding the perception of credibility thus falls on any healthcare debunker because social media platforms don't verify credentials. In this regard, TikTok has given rise to so-called ``patient influencers'' who claim no medical expertise and instead have taken to social media to share unverified advice and their experience dealing with health problems~\cite{Bushak2023}. TikTok has also given rise to a similar phenomenon of ``whitecoating'' as defined by Murtha et al., where a regular user is posing as a healthcare worker who uses their perceived authority to influence the viewer to purchase products, usually using dubious and unverified information~\cite{Murtha2022}. While professional healthcare creators on TikTok target general health-related misinformation, they also face the aforementioned subtle ways of distorting credible medical information. In this context, debunking also exposes them to harassment and bullying as it directly interferes with the ``patient influencers''' or ``whitecoaters''' economic incentives of posting content~\cite{Bautista2021}.

\section{Methodology}
Moderation efforts alone have proven ineffective at disrupting misinformation because platforms have shied away from directly debunking falsehoods circulated among users. Health professionals have taken time out of their busy schedules to engage in \lq debunk-it-yourself\rq~activities on social media, aiming to enhance the accuracy of information within their areas of expertise. We set out to investigate the workings of this novel \textit{against-misinformation} approach that shifts the responsibility from the platforms to professionals, challenging the algorithmic current on TikTok.

\subsection{Recruitment, Sampling, and Data Collection}

Before initiating our recruitment and sampling, given the nature of our study, we secured approval from the Institutional Review Board (IRB) to conduct an exploratory survey (details provided in the \hyperref[app:study-questions]{survey questionnaire} in the Appendix). We targeted health professionals who actively counter misinformation in English on TikTok without limiting participation based on the country of origin. We concentrated on nutrition and mental health, topics often exploited by \lq\lq influencers\rq\rq~with dangerous and unverified content such as self-diagnosis, treatment recommendations, weight loss advice, self-harm, and at-home recovery programs, as documented in prior studies~\cite{Harpal2023, PlushCare2022, cch2023}. Using TikTok's research API, we extracted data on these popular health-related topics recently affected by misinformation. We compiled the initial dataset with the hashtags \#adhd and \#anorexia, identifying commonly used variations that led to further misinformation on broader mental health and disordered eating issues. Additional hashtags included \#neurodivergent, \#adhdtiktok, \#wieiad, \#caloriedeficit, \#diettips, \#weightloss. To discover videos debunking misinformation, we applied the hashtags \#duet and \#stitch in these content areas.

In the recruitment process, we did not restrict ourselves by any engagement metrics of the debunking videos, and we randomly sampled a balanced set of 20 videos per topic. This was done to avoid preferential treatment to popular videos or ignore less popular ones. Each researcher individually reviewed videos to determine if they met the criteria of a debunk-it-yourself video. The research was done through credible academic journals to ensure the selected debunkers provided science-based information. We also reviewed each content creator we wanted to recruit to ensure they had credentials and/or accreditation from the relevant health bodies to provide the information they were giving. This was done by validating they worked for a health-accredited business or university that provides research or services in this area. They posted their debunking videos and verified their credentials/accreditation on the respective official websites. We also double-vetted their LinkedIn and/or Google Scholar profiles to ensure they are connected with credentialed/accredited peers who also work as health professionals in their area of expertise. 

After vetting, we contacted the participants using the email addresses provided on their LinkedIn profiles or websites to ensure we communicated with the content's original creators, as impersonation accounts are prevalent on TikTok. We utilized Qualtrics for our survey, emailing the link to $135$ debunkers. The survey, lasting on average 30 to 40 minutes, was completed by $14$ participants after we filtered out incomplete responses, responses that did not address the questions, or provided immaterial answers (total of 3 filtered our responses). Each researcher conducted this filtering individually, followed by a team consensus to remove low-quality responses. The survey responses remained anonymous, and participants could opt out of any question they were uncomfortable answering. The survey required approximately 30 minutes to complete. Participants received a \$20 Amazon eGift for for their participation, totaling \$280 for all surveys (Participants coming from countries outside the US were compensated them in with an Amazon eGift card in the local currency that amounted to \$20; all of our participants were from countries where Amazon operates an eGift card service). The demographic structure of our survey participants is detailed in Table \ref{tab:demographics}, reflecting a balanced and diverse sample.

\begin{table}[htbp]
\renewcommand{\arraystretch}{1.5}
\footnotesize
\caption{Sample Demographic Distribution}
\label{tab:demographics}
\centering
\aboverulesep=0ex % Solution part 1 of 3
   \belowrulesep=0ex % Solution part 1 of 3
\begin{tabularx}{\linewidth}{|Y|}
\hline

\textbf{Gender} \\\hline
\footnotesize
    \hfill \makecell{\textbf{Female} \\ 10 (71\%)} 
    \hfill \makecell{\textbf{Male} \\ 4 (29\%)} 
 \hfill\null
\\\hline

\textbf{Age} \\\hline
\footnotesize
    \hfill \makecell{\textbf{[21-30]} \\ 4 (29\%)} 
    \hfill \makecell{\textbf{[31-40]}\\ 8 (57\%)} 
    \hfill \makecell{\textbf{[41-50]} \\ 2 (14\%)}
\hfill\null 
\\\hline

\textbf{Political leanings} \\\hline
\footnotesize
    \hfill \makecell{\textbf{Left} \\ 8 (57\%)} 
    \hfill \makecell{\textbf{Moderate} \\ 5 (36\%)} 
    \hfill \makecell{\textbf{Apolitical} \\ 1 (7\%)} 
\hfill\null 
\\\hline

\textbf{Ethnicity} \\\hline
 \scriptsize
     \hfill \makecell{\textbf{Asian} \\ 2 (14\%)} 
     \hfill \makecell{\textbf{Black or African American}\\ 2 (14\%)} 
     \hfill \makecell{\textbf{Latinx} \\ 1 (7\%)} 
     \hfill \makecell{\textbf{White} \\ 8 (57\%)}
     \hfill \makecell{\textbf{Other} \\ 1 (7\%)} 
\hfill\null 
\\\hline

\textbf{Highest Level of Education Completed} \\\hline
\footnotesize
    \hfill \makecell{\textbf{College} \\ 8 (57\%)} 
    \hfill \makecell{\textbf{Graduate} \\ 6 (43\%)} \hfill\null 
\\\hline
\end{tabularx}
\end{table}
\raggedbottom

\subsection{Trust and Ethical Considerations}
Engaging in first-person debunking exposes debunkers to various risks, making the establishment of trust with our participants crucial. We assured them of the protections we had in place, and provided support throughout their participation. We communicated our interest in capturing the \lq\lq richness\rq\rq of their approaches and experiences throughout the `\textit{Debunk-It-Yourself}' process. We clarified that the study did not aim to assess or independently verify the quality of their debunking nor expose any direct content they created on TikTok in response to misinformation.

We assured participants that the study sought new insights and knowledge in a novel way by engaging directly with misinformation on social media at a user level. We transparently stated that the study would not require participants to engage in any new debunking activities but would focus on discussing their past efforts and experiences. To minimize risks like social desirability bias, unnecessary stress, or distress, we avoided exposure to TikTok misinformation content or related topics during the study. We maintained impartiality regarding TikTok's misinformation policies, content moderation approaches, third-party debunking services, and politicization of misinformation. We thus avoided prompting participants to disclose any attitudes, beliefs, or behaviors that could breach social or professional norms. Given that misinformation often intertwines with political issues in a scope that falls outside of this paper, we specifically steered clear of political contexts, though participants were offered the option to voluntarily disclose their political leanings if they wished as part of the survey.

After outlining the recruitment criteria as stipulated in our IRB-approved protocol, we discussed potential risks and difficulties associated with the study, particularly concerning confidentiality and the potential impact on participants’ ongoing debunking activities (such as the risk of being banned from TikTok due to platform decisions). We committed to safeguarding participants' identities, areas of expertise, and any content (debunking or otherwise) they produced on TikTok or other platforms. Given the hostile environment on social media, which often targets outspoken users, we recognized that any link between participants and a study on misinformation could endanger their online personas, subjecting them to harassment or negative psychological impacts~\cite{Schmid2023}. We chose a survey as our research instrument instead of in-person or recorded interviews to minimize any perceived risk to participants. In terms of their ongoing `\textit{Debunk-It-Yourself}' activities, we carefully reviewed TikTok’s terms of service and safety policies to ensure that our questions would not inadvertently lead to responses that could violate these policies or place participants at risk~\cite{tiktok-tos, tiktok-safety}.

Participants were informed that they could withdraw from any survey question at any time. We emphasized that we sought their perspectives to allow a broad interpretation of the sensitive attitudes, beliefs, or behaviors influencing their `\textit{Debunk-It-Yourself}' efforts without imposing any predefined expectations on their responses. At the conclusion of the survey, participants were informed that the research team would be available for up to $30$ days post-participation to help locate and remove their data if requested, though we informed them that might not be possible due to the anonymity of the responses (none requested data removal, and we were open to look up to patterns in their answers they supplied back to us to locate these records). The reason for offering the option to attempt, possibly, find, and remove any responses is to allow for the participants to reflect upon their participation, consider whether it might create any subjective risks to their safety and well-being, and have the option to mitigate these risks.

While none of the participants did feel this way, we acknowledge that the nature of the `\textit{Debunk-It-Yourself}' effort might expose individuals to risks such as harassment, threats, or various forms of abuse. Additionally, we acknowledged the possibility of misuse or misinterpretation of the study results in the broader social media discourse, which is beyond the control of the researchers (i.e., we had no safeguards against the particular risk of misinterpretation). We communicated all of these risks in the consent process and we also noted that we are understanding the complex nature of subjective assessment relative to potential exposures at the end of the interview. We did the study in the anonymous way, but to address any potential concern outside of the anonymity safeguard, we also offered the opportunity to withdraw at the end of the survey or up to $30$ days post-participation (pending locating their records successfully). We briefed participants that we were not affiliated with any social media platforms, fact-checking services, or health authorities and did not receive funding from these entities. 

\subsection{Survey Data Analysis}
Since we conducted an exploratory survey, we requested that participants provide detailed, lengthy responses to our open-ended questions. We worked with a verbose dataset as participants often provided a minimum of four informative sentences for each of their answers. We employed an inductive coding approach to analyze the collected data, identifying frequent, dominant, or significant themes within their responses. Following the guidelines suggested in~\cite{Braun2006}, we initially familiarized ourselves with the data by manually reviewing each survey response to ensure no personally identifiable information was included. Two members of the research team then conducted a round of open coding on two arbitrarily selected survey responses to capture the main stages of the `\textit{Debunk-It-Yourself}' process. Subsequent discussions about individual coding schemes led to the development of an agreed-upon basis that was used incrementally brought to a full codebook through coding the remaining survey responses.  Using this codebook, we independently coded the remaining responses, achieving an \textit{Inter-Rater Reliability} (IRR) of $k = 0.8334$ (Cohen's kappa), which was considered acceptable. The themes identified, relative to the codebook, were further discussed and interpreted, and example quotations were selected to illustrate each finding (instead of a simple summary)~\cite{fereday2006demonstrating}.

We structured a \hyperref[app:codebook]{codebook}, detailed in the Appendix, that encompasses 12 main aspects of the `\textit{Debunk-It-Yourself}' response to misinformation. The process begins with \textit{debunking initiation}, involving passive exposure to an alternative narrative on TikTok (i.e., narratives that contain incorrect facts, faulty logic, or faulty interpretations of known facts) that the provider decides warrants a direct video response. We explored the baseline our participants used to define what \textit{constitutes misinformation} on TikTok. During the \textit{debunking selection} stage, participants chose specific videos to \textit{stitch} or create duets with based on their potential to spread harmful misinformation. This stage also provided an opportunity for debunkers to report such videos to platform moderators. The next step involved selecting the \textit{DIY approach} -- the arguments, elements, and content to be included in the debunking video. Participants' approach was informed by their \textit{TikTok algorithmic familiarity}, or their understanding of the platform’s recommendation and moderation algorithms.

This understanding is closely linked with a knowledge of \textit{influencer strategies}, such as decisions made by misinformation creators to boost engagement. Participants considered these tactics while crafting their responses. With TikTok's moderation primarily based on hashtags and keywords~\cite{Ling-Gummadi}, debunkers often faced \textit{warning labels on videos}. They needed to adapt their responses accordingly, dealing with the moderation process, responding to, objecting to, or requesting clarification from TikTok's moderators. Given that multiple health professionals frequently address our selected topics, opportunities for \textit{debunking collaboration} emerged, promoting a joint effort against misinformation and peer reviews of debunking strategies. This collaboration could potentially be leveraged as an option in TikTok's efforts to incorporate the \textit{Debunk-It-Yourself} approach, provided TikTok is committed to meaningfully disrupting misinformation.

\subsection{Debunking Video Data Analysis}
In addition to the survey, respondents were given the option to provide their TikTok handle, which we could use for further analysis. Using the TikTok API, we downloaded metadata from the users' feeds and conducted a simple exploratory analysis relative to daily postings, views, likes, and hashtags used in their videos. We did this to observe whether our participants responded in kind and resembled an ``influencer'' counterpart to the one they targeted with their debunking. As our study was conducted in 2023, we decided to collect data only from that year to capture a complete annual cycle of debunking activities. We collected 26,866 videos for 135 TikTok users, of which our 14 participants were a subset, accounting for 1,649 of the videos. We collected all types of videos, out of which were mostly debunking, that our participants posted from their accounts, be that individually or as a duet stitched to another video. For the stitched videos, we left out the one that our participants targeted, though internally, we sampled, for each participant, at least five videos to ensure they indeed contained misleading claims. We undertook this approach to better illustrate the scale and impact of the \lq\lq \textit{Debunk-it-Yourself}\rq\rq~effort on TikTok, particularly in the areas of mental health and eating disorders.

\section{Results} \label{sec:results}
\begin{figure*}[h]
    \centering
    \includegraphics[width=\textwidth]{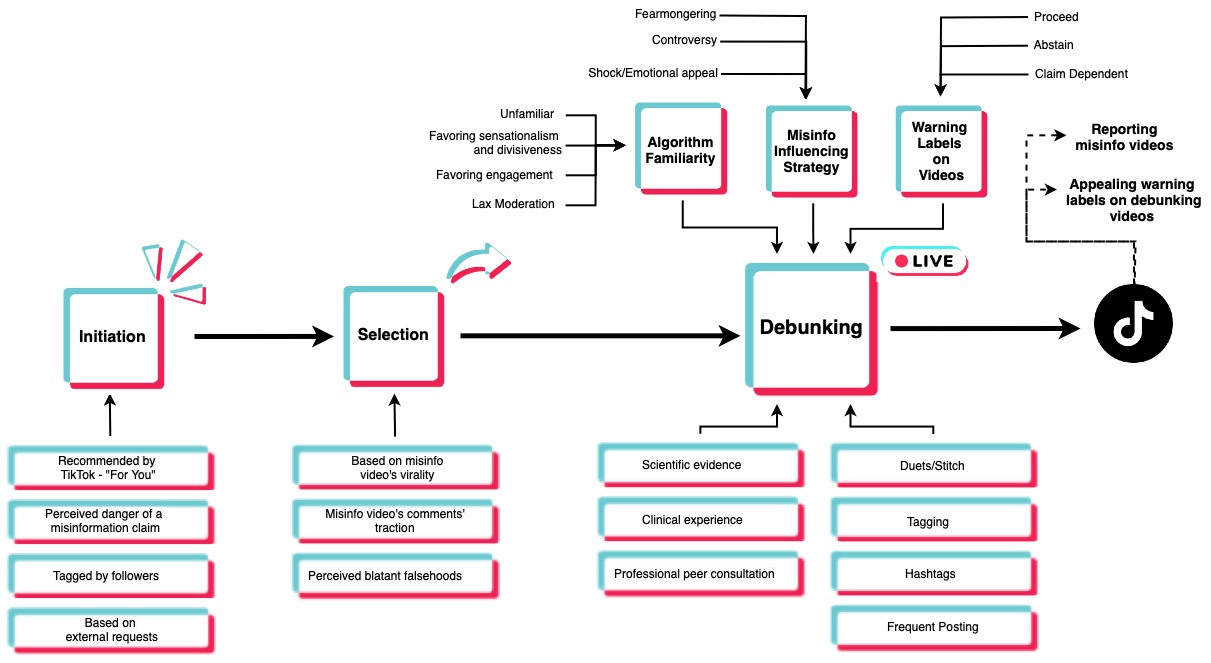}
    \caption{`\textit{Debunk-It-Yourself}' process}
    \label{fig:debunk-process}
\end{figure*}

The `\textit{Debunk-It-Yourself}' process, derived from our data, is shown in Figure \ref{fig:debunk-process}. In the following subsections, we detail each of the common stages among the sample of health professional influences in debunking falsehoods on TikTok. For each stage, we elaborate on the tactics employed to produce a debunking video that is the best fit against the target misinformation video. In addition, we include the debunkers' actions to maximize their response and their reflections on the ``\textit{Debunk-It-Yourself}'' process relative to how it might evolve in the future.

\subsection{Debunking Initiation} 
Platforms usually employ automated means to initiate a moderation action, and TikTok was found to do so, with poor results, by using hashtags and keywords in videos \cite{Ling-Gummadi}. In contrast, our participants' debunking response was initiated in four specific ways, as shown in Table \ref{tab:d-initiation}. The most prevalent initiation was simply by going on TikTok and seeing what videos the platform recommended to them, given that the algorithm tends to show topically relevant content relative to their expertise and interests \cite{tiktok-algo}. Participants didn't specifically search for misinformation -- or ``\textit{went out of [their] way to find these videos}'' -- because they ``\textit{see a lot of them on [their] FYP page}'' (\textbf{P12}). While this approach might appear too personalized and oblivious to non-recommended misinformation videos, in essence, it is similar to any platform moderation effort as both target trending health misinformation \cite{Pathak2023}. In fact, the `\textit{Debunk-It-Yourself}' is even more disruptive than the labeling or community deliberation in that it picks the most viral videos to do so, a choice that platforms might want to avoid as this drives engagement and goes against their incentives \cite{Zeng2022}. 

Next, the `\textit{Debunk-It-Yourself}' was appended by what was also recommended to the followers of our participants. Participants indicated they initiate their debunking effort when they are ``\textit{tagged in suspicious videos by people who follow [them]}'' (\textbf{P6}). This type of initiation indicates that followers trust the credibility and expertise of our participants to turn to them for debunking help. This is in contrast with the labeling and, to an extent, with the community notes as these moderation efforts are perceived as either opaque or obscure relative to credibility \cite{cose2022, Kirchner, Swire-Thompson} and lack expertise \cite{Drolsbach2023}. This was not the only debunking initiation-by-association we uncovered in the data as our participants felt responsible for addressing misinformation claims their ``\textit{friends, family, patients, and students ask about}'' (\textbf{P5}). The interesting aspect of debunking by an external request is that it does not necessarily need to come from a TikTok user, but it could be a result of any misinformation encounter. As such, and together with the follower tagging, the initiation-by-association balances out any personalized initiation that is tied to the FYP page of our participants. Equally, it allows for responding to any non-recommended misinformation videos.

Next to the passive initiation, our participants also engaged in actively seeking out target videos by assessing how dangerous a misinformation topic or set of claims could be to viewers on TikTok. Participants sought to debunk those videos, in the words of \textbf{P6} that ``\textit{get dangerously close to getting people hurt}.'' What also ``triggered'' their debunking impulses was that participants, in these cases, ``\textit{felt more compelled to respond when someone without qualifications speaks in absolutes or otherwise implies expertise}'' (\textbf{P8}). True, this adds to the personalized aspect of initiation as it adds subjectivity, and it was openly acknowledged by our participants. However, they addressed this perceived subjectivity by looking ``\textit{if a lot of other evidence-based creators [debunkers] addressing a certain video or topic to consider covering it as well}'' (\textbf{P14}).

% An altruistic approach was applied when initiating the debunking process. The respondents wanted to prevent the original posters from potentially harming people through false claims and misleading viewers.

\begin{table}[h]
\caption{Misinformation Debunking: Initiation}
\label{tab:d-initiation}
\renewcommand{\arraystretch}{1.0}
\centering
\begin{tabularx}{\linewidth}{|RcY|}
\hline
\textbf{Initiation Approach} & \textbf{\# of codes} &  \textbf{\# of videos} \\\hline
Recommended by TikTok & 8 & 862 \\
Tagged by followers & 5 & 326 \\
Perceived danger of claim & 3 & 277 \\
Based on requests & 3 & 184 \\\hline
\end{tabularx}
\end{table}

\subsection{Debunking Selection} 
TikTok's participatory nature incentivizes the frequent creation of videos, bringing ``influencers'' who peddle unverified and dangerous content relative to nutrition and mental health to the fore \cite{Harpal2023, PlushCare2022, cch2023}. As the abundance of such content could be overwhelming, our participants noted that they usually had to select what videos to target after the debunking was initiated. The selection, as shown in Table \ref{tab:d-selection}, was based on three distinct approaches. Most of our participants selected the target misinformation videos based on their virality, i.e., those with a high view count and the content creators have a large following audience on TikTok. Participant \textbf{P2} followed a simple rule, for example, looking at ``\textit{how much of a following the original creator has -- the more followers, the more influence, potentially the more damage they can do)}.'' 

\begin{table}[htbp]
\caption{Misinformation Debunking: Selection}
\label{tab:d-selection}
\renewcommand{\arraystretch}{1.0}
\centering
\begin{tabularx}{\linewidth}{|RcY|}
\hline
\textbf{Selection Approach} & \textbf{\# of codes} &  \textbf{\# of videos} \\\hline
Virality & 8 & 608 \\
Blatant Falsehoods & 7 & 679 \\
Susceptibility & 4 & 362 \\\hline
\end{tabularx}
\end{table}

The view count and number of followers are also used in both the labeling and the community notes moderation \cite{MartelRand2023, pilarski2023community}. However, the crucial difference lies in response to the associated misinformation content -- the labels and the community notes are less viral \cite{Drolsbach2023}, but a `\textit{Debunk-It-Yourself}' video has no such a constrain. As the participants in our study also fit the ``influencer'' profile (more on their content creation behavior in Section \ref{sec:debunking-behaviour}), their debunking content has the potential to also reach the level of virality as the targeted misinformation videos. This response in kind is unique and departs from the calculated platform interventions, which, in our view, demonstrates the first materialization of the prescribed actions in the \textit{Debunking Handbook} \cite{Lewandowsky2020}. 

Aside from the virality, another selection criterion employed by our participants was how blatant the falsity of a misinformation video was. This assessment was done by the participants themselves, relative to the existence of directly disproving information as well as the ``\textit{direct potential harmfulness}'' of the claims in the video (\textbf{P5}). This criterion is central to the moderation effort and the community notes as their main goal is to share timely information or credible content from third-party sources for well-established historical, scientific, and health topics that are often subject to misinformation~\cite{youtube-misinfo, meta-misinfo, twitter-misinfo, tiktok-safety}. The `\textit{Debunk-It-Yourself}' effort goes beyond this ``timely information sharing'' as the debunkers (i) don't usually turn to third-party sources; (ii) don't send users outside of the platform; (iii) offer a user the well-established scientific and health facts through content alongside the misinformation, not as a platform affordance beneath it (more on this in Section \ref{sec:debunking-approach}). 

Another reason why the `\textit{Debunk-It-Yourself}' effort goes beyond platform moderation is that participants also looked further into the impact the video was already having on viewers. As participant \textbf{P7} noted, they look at the engagement and the comments on the video as a selection criterion -- ``\textit{the more people that agree or comment things such as `I didn't know this' claim, the more likely [they] will select the video to respond to it}.'' Our participants, therefore, selected videos for debunking that were ``\textit{something in alignment with the areas of expertise [they] feel comfortable speaking about}'' (\textbf{P13}). The unique and new thing here is that participants utilize their expertise in their field to quickly identify misinformation and determine its impact without waiting for it to become viral in the first place. This is an important step towards a comprehensive disruption of misinformation because it addresses both the falsity of the video and the users' susceptibility to it, i.e., exposing the wrongness of the claim and dissuading the users against it.

Participants in our study also considered the target audience they wanted to reach out, given that the TikTok user base is highly diverse relative to age, education, as well as digital and media literacy. Aiming to reach as wide an audience as possible, our participants factored the need to include ``\textit{captions so not just young people  could see it but also more senior users to be able to get the message}'' (\textbf{P2}). Target audience that could be particularly susceptible to falsehoods related to nutrition or mental health self-diagnosis, such as young and impressionable users, were particularly considered by our participants, as ``\textit{the threat of harm is more imminent}'' (\textbf{P7}), so they opted to use ``\textit{ trending, non-threatening vernacular that could easily get through them}'' (\textbf{P14}). Participants were also aware that there are experienced people who are drawn to misinformative content but lack literacy skills to dismiss it, so our participants also considered ``\textit{conveying the debunking in simple enough therms enough for these users to clock [the original video] as misinformation.}'' (\textbf{P2}).

\subsection{Debunking Approach} \label{sec:debunking-approach}
At the heart of the platforms' moderation approach are third-parties or ``fact checkers.'' The fact checking process displaces the burden of truth arbitration from platforms to individual services that, evidence shows, are subject to bias, lack of expertise, and selectiveness \cite{Walter2020}. Even the community notes an approach is a form of a ``third-party fact checker'' as non-expert users serve as truth arbiters. In stark contrast to this, our participants did not simply check facts as a binary outcome of a true/false exercise. Instead, our participants followed the prescribed \textit{Debunking Handbook} approach to the letter -- they first explained why one narrative is false or misleading in their videos, and second, what is the actual truth instead \cite{Lewandowsky2020}.

To do so, our participants, as shown in Table \ref{tab:d-approach}, relied on scientific evidence, often in combination with their clinical experience and consultation with their peer professionals. Our participants were cautious about utilizing their personal experience as it, in their view, could potentially be perceived as bias -- a trait that is usually associated with platform-led moderation. Participant \textbf{P3} nicely summed this consideration: 

\begin{quote}
``\textit{I would never debunk based off of just my experience because that is just bias against bias. If I ever share my experience, I will name it as exactly that, my experience. I look to academic journals and research from respected places and always check conflicts of interest}.''    
\end{quote}

Providing accurate information was critical to the participants in that they ``\textit{consult with peers offline before finalizing the content of the video, in addition to referencing academic resources}'' (\textbf{P8}). Transparency was a trait that was also important to our participants as they went as far as to ``\textit{cite [their] sources clearly by showing them in the background behind [them] using the green screen effect}'' (\textbf{P6}) in their videos. Interestingly, only a couple of participants mentioned that they sometimes utilize fact checking services. Contrary to moderation efforts resting entirely on fact-checking, most of our participants explicitly distanced themselves from any ``third party.'' For example, participant \textbf{P12} stated that they debunk misinformation videos on TikTok by using ``\textit{academic journals, studies, and stats from credible sources and organizations, but never using fact-checking services}'' (\textbf{12}). The reason for this, in the participant's view, was to remove any association with external efforts that could potentially damage their reputation, both on TikTok and professionally.

\begin{table}[htbp]
\caption{Misinformation Debunking: Approach}
\label{tab:d-approach}
\renewcommand{\arraystretch}{1.0}
\centering
\begin{tabularx}{\linewidth}{|RcY|}
\hline
\textbf{Debunking Approach} & \textbf{\# of codes} &  \textbf{\# of videos} \\\hline
Scientific evidence and clinical experience & 8 & 1023 \\
Scientific evidence alone & 4 & 487 \\
Scientific evidence and consultation with peers & 2 & 139 \\\hline
\end{tabularx}
\end{table}

\subsection{Debunking Video Creation} \label{sec:debunking-video-creation}
With a target video selected and a determined debunking approach, our participants shared how they actually did the video-to-video response on TikTok. As shown in Table \ref{tab:debunking-video-creation}, our participants attempted to use the virality of the original video to get their debunking videos to gain viewers through either: (i) duetting or stitching the original video, (ii) tagging the original creator; (iii) or using the same hashtags as the original video in hopes it will show on the same users' FYP feeds. Each of these strategies has its own pros and cons, according to our participants, and they decide based on the target video's particular claim. When they go for duetting, participants stated they ``\textit{appear as interesting, polite, and clear and usually stitch it (play their video for the first 5 seconds) and then respond by a catchy hook to capture interest and try to maintain that attention throughout the entire video}'' (\textbf{P6}). 

While this might help with the debunking video gain view count, it was of concern to our participants because ``\textit{it would [also] send traffic to the spreader of misinformation}'' (\textbf{P5}) and increase the view count of their video too. But the threat of direct mutual amplification was not entirely dissuading on our participants. Aware of it, participant \textbf{P8} said that in addition to the duetting, they:

\begin{quote}
``\textit{...would use the same hashtags or include a call-to-action in the caption or video to encourage my audience to help me set the record straight. At times, I'll even include a snarky statement like, `I hope you make this go as viral as the original.'}''
\end{quote}

Other participants opted out of tagging the video creator instead and used ``\textit{clips of their video} (\textbf{P5}) as points for debunking. A main reason for staying away from duetting, according to our participants, is their direct exposure to ``\textit{retaliation or direct harassment on TikTok}'' (\textbf{12}). For example, participant \textbf{P3} mentioned that they have ``\textit{gotten absolutely attacked before for duetting}.''

\begin{table}[!h]
\caption{Misinformation Debunking: Own Influencing Tactics}
\label{tab:debunking-video-creation}
\renewcommand{\arraystretch}{1.0}
\centering
\begin{tabularx}{\linewidth}{|RcY|}
\hline
\textbf{Debunking Video} & \textbf{\# of codes} &  \textbf{\# of videos} \\\hline
Duets/Stitch & 9 & 1184 \\
Tagging & 3 & 364 \\
Hashtags & 2 & 101 \\\hline
\end{tabularx}
\end{table}
% \vspace{-2em}

\subsection{Debunking Against the Algorithm}
Our participants were aware that simply duetting or tagging a video won't in and of itself help the debunking get across the TikTok audience. Though most of them were hardly familiary with the recommendation algorithm, as shown in Table \ref{tab:d-familiarity}, they conjectured it favored engagement and sensationalism without much moderation or fairness. Some of the participants understood the basic engagement approach that ``\textit{platforms give recommendations based on interests of the user and this is based on what they click on, what they watch, what they search for, what they save and what they comment on}'' (\textbf{13}), but they felt that it was working against them. Participant \textbf{P7} shared that ``\textit{the algorithms on TikTok tend to promote videos based off of engagement, so people who spread misinformation tend to get pushed more because more people engage with the content}'' and participant \textbf{P7} believed ``\textit{algorithms tend to favor sensationalist and divisive content vs. accurate information}.''

The feeling that they are fighting an uphill battle is not entirely new sentiment considering that the labeling, in particular, was also perceived as unfair \cite{Swire-Thompson, cose2022}, selective or lax \cite{Kirchner}, and the selective ``shadowbanning'' of users \cite{Johns2024}. In the case of ``\textit{Debunk-it-Yourself},'' the preference against the debunkers' videos is not tied to a misinformation policy as is the case with the platform moderation, but the recommendation algorithm itself, making it harder to append and appeal. Our participants saw the magnitude of this hurdle is further highlighted and appended by the ``stickiness of misinformation,'' as participant \textbf{P11} put it:  ``\textit{The algorithms are set up to show more of what you enjoy and interact with, which creates a cognitive bias as people end up seeing more and more of a misinformation they may want to believe thus reinforcing it in their minds}.''

The lax moderation didn't help either, with participants expressing that TikTok ``\textit{will rarely remove videos which are false}'' (\textbf{7}). A sense of ``shadowbanning'' was also present among our participants, who felt that TikTok was working against their efforts to fight misinformation. In the words of participant \textbf{P12} ``\textit{[TikTok] doesn't do a good job of removing or labeling misinformation -- if anything, they tend to remove the things that help people share the truth and keep the more harmful things on there}.''  Even if this is hard to prove, our participants saw the overall moderation effort on TikTok as non-conductive to truth arbitration, even if the users are interested in joining this effort, in the words of participant \textbf{P16}: 

\begin{quote}
``\textit{My understanding is that platforms have evolved in recent years to include more robust fact-checking features which, may be from within the platform or initiated and crowd-sourced from users. There are reporting features to flag misinformation, which I have used frequently. However, I'm not sure the speed or efficacy at which these features work or to what degree content needs to be flagged to declare it misinformation. I almost never have success with reporting a video with blatant misinformation and having it removed or deleted.}''     
\end{quote}

\begin{table}[htbp]
\caption{Misinformation Debunking: Algorithm Familiarity}
\label{tab:d-familiarity}
\renewcommand{\arraystretch}{1.0}
\centering
\begin{tabularx}{\linewidth}{|RY|}
\hline
\textbf{Algorithm Familiarity} & \textbf{\# of codes} \\\hline
No in-depth familiarity & 10 \\\
Favoring Engagement & 3  \\\
Lax Moderation & 3 \\\
Favoring Sensationalism & 1 \\\hline
\end{tabularx}
\end{table}

\subsection{Debunking Against Virality}
Misinformation gains traction easily on social media platforms, and our participants -- based on the extensive exposure -- were confident that shock and emotional appeal were primary drivers of success for a misinformation video to spread, as shown in Table \ref{tab:d-virality-factors}. In the words of participant \textbf{P6}, ``\textit{the truth is hardly ever as interesting compared to ideas and concepts that are extremely compelling, emotional, and relatable}.'' In addition to the emotional appeal, and in the context of the topics of eating and mental health disorders, participants felt that fear in particular ``\textit{is a stronger emotion than enlightenment}'' (\textbf{4}) and misinformation influencers do on TikTok do a great deal of fearmongering. Controversy was another factor harnessed towards the virality of misinformation videos on TikTok. Participants felt that the participatory nature of the platform is ideal to fit videos on ``\textit{topics that are familiar enough to the viewer so they know they should care about them, but not familiar enough that they're knowledgeable enough to clock it as misinformation}'' (\textbf{8}).

Moderating against virality, both in the case of labeling and community notes, also considers these factors; however, they are largely asymmetric. The mere exposure to a misinformation video generates a strong and automatic affective response, but the warning may not generate a response of an equal and opposite magnitude \cite{Gawronski}. This is because the labels often lack meaning, have ambiguous wording, or ask users to find context themselves, which is cognitively demanding and time-consuming \cite{Butler2023}. The ``\textit{Debunk-It-Yourself}'' effort precisely addresses this drawback in a symmetrical way -- our participants shared that they create debunking videos of equal affective magnitude that include context, meaning, and references, use simple and relatable wording, and make far less of a cognitive demand for users to dismiss a stitched or tagged misinformation video.

\begin{table}[htbp]
\caption{Misinformation Debunking: Virality Factors}
\label{tab:d-virality-factors}
\renewcommand{\arraystretch}{1.0}
\centering
\begin{tabularx}{\linewidth}{|RY|}
\hline
\textbf{Virality Factors} & \textbf{\# of codes} \\\hline
Shock/emotional appeal & 9 \\\
Fear mongering & 4  \\\
Controversy & 4 \\\hline
\end{tabularx}
\end{table}

\subsection{Debunking in Presence of Moderation}
Though TikTok applies coarse moderation \cite{Ling-Gummadi}, warning labels frequently appear on health-related videos that peddle misleading narratives \cite{sharevski2023abortion}. For our participants, the decision to debunk misinformation was largely unimpacted by TikTok's warning labels because ``\textit{it can still be worth talking about why the claim is dangerous}'' (\textbf{P9}). In fact, some of them said that a label would ``\textit{make [them] more likely to duet}'' (\textbf{P7}) the misinformation video with a debunking one of theirs. Participants also saw the opportunity to demonstrate why debunking makes a better case against misinformation. As participant \textbf{P8} put it,``\textit{[their] audience may have not watched [the labeled video] and it's still valid to share}'' it with a debunking video counterpart. Some participants felt their decision depended on the content. For example, participant \textbf{P9} explained that ``\textit{if it's obvious that it's misinformation (i.e., the claims are so ridiculous that nobody would ever believe them) then [they] are typically satisfied by a label, but if the claims are convincing and the response in the comments seems robust, [they] will debunk it} (\textbf{9}). Only one participant, as shown in Table \ref{tab:d-labeled}, was satisfied by a label in case they encountered one during the initiation and selection phases.

\begin{table}[htbp]
\caption{Misinformation Debunking: Warning Labels}
\label{tab:d-labeled}
\renewcommand{\arraystretch}{1.0}
\centering
\begin{tabularx}{\linewidth}{|RY|}
\hline
\textbf{Impact of warning labels} & \textbf{\# of codes} \\\hline
Proceed with debunking & 5 \\\
Claim dependent & 4  \\\
Hasn't seen warning labels & 4 \\\
Abstain from debunking & 1 \\\hline
\end{tabularx}
\end{table}

\subsection{Post-Debunking Response}
After posting a debunking video, our participants shared that they must take additional work to complete the entire ``\textit{Debunk-It-Yourself}'' process. This includes deciding if they will report the target misinformation video and monitoring the reception of the debunking video. As shown in Table \ref{tab:d-reporting}, participants were split on whether to report the original video as misinformation to TikTok or not. The ones that were open to report shared that they would do it because ``\textit{thought were extremely harmful}'' and, as a result, have ``\textit{successfully had them taken down}'' (\textbf{5}). Those that were not, justified the lack of follow-up action as the right for any creator to ``\textit{share [what they deem important] in their views and opinions},'' which will allow misinformation to surface in the first place and get debunked appropriately, according to \textbf{P13}. Another reason to avoid reporting was the feeling that TikTok wouldn't do much about the video. In the experience of participant \textbf{P8}, there is ``\textit{almost never a success with reporting a video with blatant misinformation and having it removed or deleted}.'' Some of the participants, like \textbf{P10}, noted they report misinformation videos ``\textit{sometimes, not always, depending on how dangerous the video is}'' (\textbf{P10}).

\begin{table}[htbp]
\caption{Misinformation Debunking: Reporting}
\label{tab:d-reporting}
\renewcommand{\arraystretch}{1.0}
\centering
\begin{tabularx}{\linewidth}{|RY|}
\hline
\textbf{Reporting Misinformation} & \textbf{\# of codes} \\\hline
Yes & 6 \\
No & 6  \\
Claim dependent & 2 \\\hline
\end{tabularx}

\end{table}

As the ``\textit{Do-it-Yourself}'' process is not anonymous like the labeling or the community notes, our participants shared that they need to monitor the comments on their videos in case they need to refute any contradictory information further. They also shared that they need to ensure that TikTok has not incorrectly moderated their video. Though many participants haven't experienced direct moderation, as shown in Table \ref{tab:d-moderated}, there were participants whom TikTok had wrongly moderated. For example, participant \textbf{P9} shared that TikTok removed their debunking videos, but ``\textit{once [they] submitted the appeal and it is reviewed, the videos have been reinstated almost instantly}.'' Participant \textbf{P10} appealed for having their videos moderated with warning labels, ``\textit{asking [TikTok] to be re-looked, after which, the warning label was removed}.'' One participant, \textbf{P8}, felt they were ``shadowbanned'' by TikTok because ``{the algorithm has, at times, only pushed [their] content out to followers instead of through hashtags, audio, or other places where it could organically show up for non-followers}.'' 

\begin{table}[htbp]
\caption{Misinformation Debunking: Being Moderated}
\label{tab:d-moderated}
\renewcommand{\arraystretch}{1.0}
\centering
\begin{tabularx}{\linewidth}{|RY|}
\hline
\textbf{Being Moderated} & \textbf{\# of codes} \\\hline
No & 9 \\
Wrongly moderated & 4  \\
Shadow-banned & 1 \\\hline
\end{tabularx}

\end{table}

\subsection{Reflections on ``Debunk-It-Yourself''}
As the `\textit{Debunk-It-Yourself}' is a relatively new and unexplored phenomenon, we asked our participants about their perspectives on potential collaboration between them or others interested in debunking misinformation on TikTok. As shown in Table \ref{tab:d-collaboration}, most participants were keen on doing the debunking individually though they most stated they admire other debunkers on the platform because they, like them, use ``\textit{a non-biased approach to information, backed by scientific arguments}'' (\textbf{13}), as well as ``\textit{deliver the factual information compassionately and leaves room for a nuanced conversation}'' (\textbf{10}). Indirectly, as participant \textbf{P5} pointed out, ``\textit{debunkers have their own way of doing things}'' and following each other suffices for a distributed response against misinformation creators. However, the collaboration was not entirely off the table as some participants, like \textbf{P11}, ``\textit{created a chat room where [they] gathered some 50 debunkers to talk, share, help each other because everyone has different experiences, opinions, ways to debunking}.''

\begin{table}[htbp]
\caption{Debunking Collaboration}
\label{tab:d-collaboration}
\renewcommand{\arraystretch}{1.0}
\centering
\begin{tabularx}{\linewidth}{|RY|}
\hline
\textbf{Collaborative Effort} & \textbf{\# of codes} \\\hline
Indirect collaboration & 9 \\
Direct collaboration & 3  \\
No collaboration & 2 \\\hline
\end{tabularx}

\end{table}

We also asked our participants how TikTok could leverage their content to respond better and curb dangerous health misinformation. Most of them, as shown in Table \ref{tab:d-leverage}, felt that TikTok ``\textit{could tailor the algorithm to give evidenced-based accounts and videos (like [theirs]) more reach on the platform}'' (\textbf{P14}). Quoting the \textit{Disinformation Handbook} directly, participant \textbf{P6} mentioned that TikTok should focus on ``\textit{prebunking or promoting [their] videos so that people see them `before' they see the harmful content}. Wary of the specific participatory nature, participant \textbf{P5} added that TikTok ``\textit{absolutely needs to prioritize allowing our followers to actually see [their] content, because that's the only way the platform could help}.'' Some of the participants felt they needed to be compensated so they could respond in kind with the target influencers and ``\textit{afford to pay for editors or animators to do fun educational things with my videos}'' (\textbf{P11}). Others, like participant \textbf{P2}, felt that TikTok anyhow doesn't ``\textit{pay too much attention to misinformation}'' so any leveraging of the debunking would be counterproductive to them as their ``\textit{focus is more on getting users to spend time on their app, instead}.''

\begin{table}[htbp]
\caption{Leveraging Debunking}
\label{tab:d-leverage}
\renewcommand{\arraystretch}{1.0}
\centering
\begin{tabularx}{\linewidth}{|RY|}
\hline
\textbf{Opportunity to leverage} & \textbf{\# of codes} \\\hline
Prioritize their content & 6 \\
No leveraging & 5  \\
Compensate them & 3 \\\hline
\end{tabularx}
\end{table}

\subsection{Debunking Behavior} \label{sec:debunking-behaviour}
\begin{figure}[htbp]
    \centering
    \includegraphics[width=0.97\columnwidth]{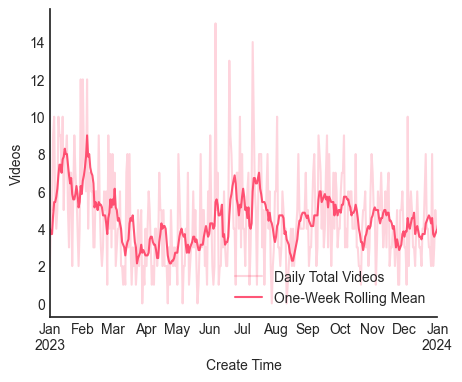}
    \caption{Total number of videos posted each day.}
    \label{fig:all_videos}
\end{figure}

To explore the `\textit{Debunk-It-Yourself}' paradigm from a quantitative perspective, we first set to plot the distribution of the frequency of posting videos by our participants, most of which are debunking in nature (we used their TikTok handles as there were public and all of the participants were willing to provide them). Though we had a relatively small number of participants, our preliminary analysis (and vetting) showed that all of them are very active on the platform (i.e., we didn't have one or two individual debunkers to stand out and skew the analysis). As shown in Figure ~\ref{fig:all_videos}, our participants were relatively consistent in posting daily and weekly throughout 2023. These trends show that our study sample aims to meet the ``influencer'' objective in frequent debunking and responding in kind for each misinformation narrative for both the nutrition and mental health topics. Anecdotally, perhaps, we noticed a peak of nutrition and mental health debunking in the first month of the year, which might be related to New Year's weight loss resolutions (nutrition) and January blues (mental health) periods, respectively.

%Interestingly, the top 10 health professional content creators from our population (\textit{n=135}) made 10,786 videos or 40.2\% of the total 26,866 videos. While there is some fluctuation at different points in 2023 among the bottom 94 posters, the top 10 posters consistently post a pretty steady number of debunking videos throughout the year. This trend is highlighted in Figure \ref{fig:top_10}. These health professional ``influences'' were balanced per topic, i.e. nutrition and mental health. Upon further inspection of their videos, we noticed that all of them employ the ``influencer'' formatting, that is, they are created in a second-person view \cite{Cheng2024}, had mostly positive titles \cite{Chen2021}, and express positive emotions \cite{Li2021}. 

%\begin{figure}[h]
%    \centering
%    \includegraphics[width=\columnwidth]{TikTok Misinformation Debunkers/graphs/all_videos_top10.png}
%    \caption{Total number of videos posted each day split by the 10 debunkers who posted the most against the remaining 120 from our population. }
%    \label{fig:top_10}
%\end{figure}

\begin{figure}[htbp]
    \centering
    \includegraphics[width=0.97\columnwidth]{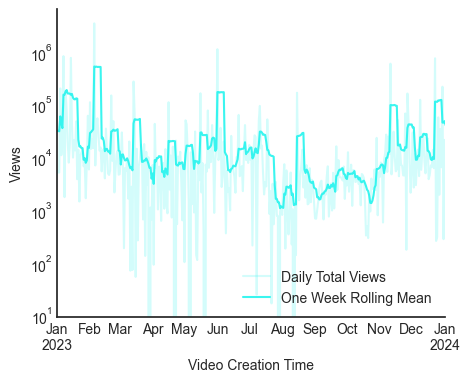}
    \caption{Total number of views of debunking videos.}
    \label{fig:total_views}
\end{figure}

\begin{figure}[h]
    \centering
    \includegraphics[width=0.97\columnwidth]{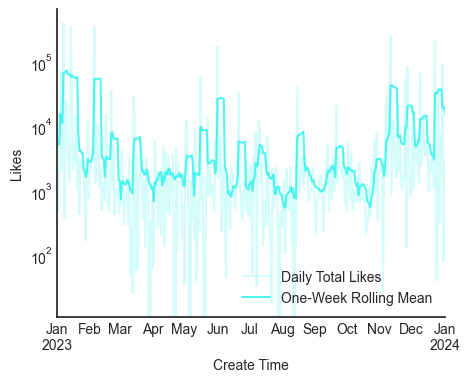}
    \caption{Total number of likes of debunking videos.}
    \label{fig:total_likes}
\end{figure}

Our analysis shows that debunkers in our population, true to their ``influencer'' posture, indeed attract an enormous engagement relative to views and likes on their videos, as shown in Figure ~\ref{fig:total_views} and Figure ~\ref{fig:total_likes}, respectively. The time of the year, perhaps related to creating and posting debunking videos in response to misinformation, also shows as a factor here. We noticed a relative lull in the total number of views from August until November, with a slight increase in December and January. Anecdotally, again, we suspect that the reason behind this viewing behavior -- which is also affected by TikTok's recommendation algorithm -- might be the seasonal cycle of trending viewership interests relative to nutrition and mental health. 

\begin{figure}[htbp]
    \centering
    \includegraphics[width=0.97\columnwidth]{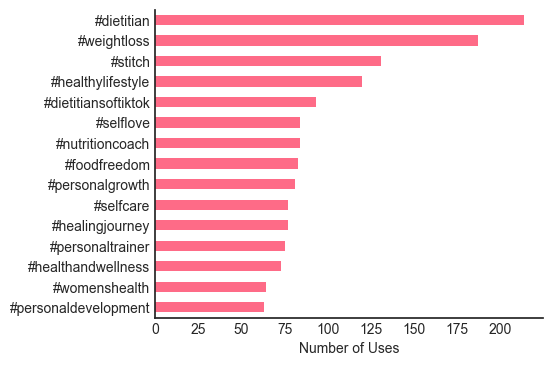}
    \caption{Distribution of hashtags used in debunking videos.}
    \label{fig:hashtags}
\end{figure}

We also looked at the hashtags that the debunking population used as another element of the ``influencing'' on TikTok. Though their relevance is constrained, considering that the participation on the platform is centered around viewing videos suggested by a central recommendation algorithm, hashtags still bear relevance as input in this suggestion \cite{tiktok-algo}. The population of debunkers used 2,481 total hashtags, though only 61 were unique. The top 15 most used hashtags are shown in Figure \ref{fig:hashtags}. Of the top five hashtags, four are related to weight loss (\#dietitian, \#weightloss, \#healthylifestyle, and \#dietitiansoftiktok). The third hashtag is \#stitch, indicating the main affordance, or feature, that the debunkers utilize to disrupt TikTok's misinformation videos.

\section{Discussion}
The abundance of questionable nutrition and health-related videos on TikTok under circumstances of lax moderation \cite{Ling-Gummadi} forced individual users who are health professionals themselves to devise their own response -- create a \textit{counter-misinformation} video against ones they believed needed immediate debunking to prevent harm. This was mostly done by selecting the videos with the highest view count and engagement, however, the discussion these videos provoked in the comments was also considered to anticipate its harmful influence. Our findings indicate that with the target misinformation video selected, the debunking videos were strongly based on scientific evidence. A unique feature we uncovered here is that this segment of TikTok users avoided \textit{using external fact checking services} -- the essence of the platform response against misinformation -- but instead rely on science, their clinical experience, and consultation with their peers. The debunking videos -- first pointing out the falsity of a claim and then offering the related scientific evidence -- were created and ``stitched'' directly to the targeted misinformation video, or they tagged its creators. This is another unique feature of the `\textit{Debunk-It-Yourself}' model in that it offers a \textit{symmetric} response elicited by both ``dueted'' videos. The platform-lead moderation is usually asymmetric in providing warning labels underneath the content that are often ignored, confusing, or asks viewers to determine the falsity themselves. 

The `\textit{Debunk-It-Yourself}' went against TikTok's algorithmic curation and moderation mechanisms, which often prioritize sensational and controversial content. These circumstances forced a debunking strategy of frequent posting, which targets viral misinformation videos characterized by sensationalism, controversy, and fearmongering. Despite the risk of \lq\lq shadowbanning,\rq\rq~a term used to describe the reduction in content visibility without the user's knowledge, the TikTok users we surveyed continued their debunking efforts. They were not deterred by warning labels TikTok might place on misinformation videos, believing these labels were insufficient in mitigating the spread of false information. Additionally, they expressed skepticism about the effectiveness of reporting videos through TikTok's standard mechanisms. In several instances, their own videos, created to debunk misinformation, were mistakenly tagged with misinformation labels. 

This led to vigorous protests, eventually resulting in TikTok removing these erroneous labels. This experience underscored the perception that TikTok's moderation system might not fully appreciate or support individual counter-misinformation efforts. The overarching sentiment that emerged from our study was one of disillusionment; they felt their contributions were at odds with the platform's primary goal of maximizing viewer engagement. Despite their status as \lq\lq influencers\rq\rq~in the realm of counter-misinformation, there was a pervasive sense that this users' work was marginalized, reflecting a disconnect between the platform's operational priorities and the ethical implications of spreading accurate information. This highlights a critical tension within social media ecosystems, where the drive for high engagement often conflicts with the imperative to foster informed and healthy public discourse involving trustworthy information.

\subsection{Threats to `\textit{Debunk-It-Yourself}'}
A significant concern regarding the `\textit{Debunk-It-Yourself}' initiative is its potential misuse or exploitation for malicious purposes. Although the platform's reliance on third-party or user consensus moderation fosters transparency and helps establish objectivity in countering misinformation, this approach lacks a robust mechanism to ensure the integrity of individual debunkers. These individuals are neither obliged to disclose their motivations nor held accountable for their objectivity, except perhaps by other health professionals on TikTok. Without effective checks and balances, there is a real risk that a debunker might target another user for personal reasons or begin spreading misinformation themselves. TikTok's inconsistent track record in handling misinformation does not instill confidence that it can effectively mitigate the risks posed by a \textit{rogue debunker}. A possible alternative could involve professional medical associations imposing standards on these individuals, which would necessitate content review and verification linked to their credentials. However, this approach raises its own set of challenges. Addressing these vulnerabilities is crucial for the `\textit{Debunk-It-Yourself}' model to gain and maintain trust among TikTok users and the broader social media community.

\subsection{Implementation Challenges of `\textit{Debunk-It-Yourself}'}
The `\textit{Debunk-It-Yourself}' initiative is primarily driven by health professionals who voluntarily engage in debunking misinformation about nutrition and mental health. Our analysis reveals that these professionals invest considerable time in creating and sharing debunking videos, often in addition to their regular job responsibilities. Although they are incentivized by TikTok's creator rewards program~\cite{tiktok-creator-rewards}, any changes to this program could significantly impact their continued participation. It is conceivable that TikTok could adjust its rewards scheme to specifically support debunking health misinformation, but such a move might conflict with the platform's commercial objectives. Furthermore, alternative forms of recognition from medical associations or employers seem impractical and unlikely to motivate widespread participation. 

Even with a provisional change in the incentives for debunking and recognition from medical associations, the `\textit{Debunk-It-Yourself}' effort is essentially at odds with the network effects that TikTok enjoys from promoting content that invokes more of an emotional, rather than rational, response characteristic for the debunking content produced by the participants in our study \cite{anderson2014privacy}. Without an internal platform regulation or intervention to promote (instead of demote or ``shadowban'') individual debunkers, the race between them and the content creators (or ``influencers'') they counter might be lost due to the bigger network power the later ones enjoy as a more salient profit makers for TikTok. Without an external governance or regulation to append truthfulness on social media (which, granted, might be one of the hardest problems we as society currently face), TikTok's successful evasion of truth arbitration and dismissal of the `\textit{Debunk-It-Yourself}' effort will be, or perhaps its already, a tempting model for other social media companies such as Instagram/Facebook or YouTube to do the same to individual debunking content that might have been created on their platform or might have been transferred from TikTok onto their platforms.

The feasibility of scaling the `\textit{Debunk-It-Yourself}' process is also a concern, particularly as misinformation is not limited to just nutrition and mental health but extends to other critical areas like vaccinations. Engaging a sufficient number of health professionals to cover the broad spectrum of health misinformation presents a formidable challenge. Increased participation could lead to conflicts over the accuracy of debunking efforts, potentially resulting in segmentation or polarization—effects that misinformation seeks to achieve~\cite{Lewandowsky2020}. Additionally, expanding `\textit{Debunk-It-Yourself}' to other platforms like Instagram, YouTube, Facebook, or Twitter would depend on the availability of similar interactive features to TikTok's stitching or duetting. While such expansion could potentially reach a wider audience, each platform's unique moderation strategies might limit the effectiveness of independent debunking efforts. Alternatively, establishing a standalone repository for debunked content could align with traditional fact-checking services like FactCheck.org or Snopes, but this approach might not be scalable, attractive to debunkers, or well-received by social media users.

\subsection{Recommendations from Participants on `\textit{Debunk-It-Yourself}'}
Our study elucidates several recommendations from participants that aim to augment the effectiveness of the `\textit{Debunk-It-Yourself}' process on TikTok. These recommendations underscore the need for both systemic changes at the platform level and refined strategies at the individual level to bolster the counter-misinformation response. Firstly, participants urge platforms to refine algorithmic support for debunking activities by strategically prioritizing content that effectively counters widespread misinformation. Additionally, there is an obvious need to modify moderation systems to minimize the erroneous flagging of debunking videos as misinformation. This adjustment will protect well-intentioned debunkers from undue penalization and support their efforts in combating falsehoods. Moreover, there is a strong advocacy for increased transparency in the operational algorithms of social media platforms. Participants express the need for platforms to disclose more about the criteria used for promoting or demoting videos, which in turn, will provide users with a insight into the mechanics behind content visibility and foster a more trustful relationship between users and platforms.

To ensure the credibility of content creators, participants propose the implementation of robust verification processes. Such measures would distinguish qualified health professionals from unqualified sources that propagate health misinformation, thereby enhancing the reliability of health information across social media platforms. In terms of collaboration, participants highlight the potential benefits of enhanced tools on social media platforms that would enable more effective partnerships among health professionals. These tools could facilitate the co-creation of content and mutual support in debunking efforts, making the overall process more efficient and widespread. Furthermore, there is a recommendation for platforms to provide educational resources or training focused on the best practices for debunking. By equipping new debunkers with effective communication and engagement strategies, platforms can expand the community of informed content creators committed to upholding factual accuracy. Lastly, participants recommend the creation of a more supportive community for debunkers through platform features that allow users to easily follow and receive updates from trusted health professionals and seasoned debunkers. Such community-building features would not only support debunkers but also enhance the public's ability to access reliable information readily.

\subsection{The Future of `\textit{Debunk-It-Yourself}'}
The `\textit{Debunk-It-Yourself}' initiative, currently focused on combating health misinformation on TikTok, shows potential for expansion into other areas of misinformation, particularly political misinformation. This type of misinformation is rapidly gaining attention on TikTok, utilizing interactive formats such as duetting~\cite{Medina-Serrano, Bandy}. The adoption of a direct debunking approach by political scientists or public officials, however, remains to be seen due to the highly contentious nature of online political discourse. Additionally, public health crises, like pandemics, offer a critical arena for the application of direct debunking strategies by health authorities. Historical trends indicate a preference for less confrontational approaches, such as passive myth-busting strategies~\cite{CDC-Myth}, but the urgency and widespread impact of misinformation during such crises could necessitate more direct interventions.

The `\textit{Debunk-It-Yourself}' approach as a supply side of the debunking efforts entails a consideration of misinformation from a national security perspective ~\cite{anderson2014privacy}. A social media platform shrouded in misinformation with far more users and engagement is a much larger of an intelligence asset than a platform with less users and activity due to the chilling effects of the individual debunking (and for that matter, any direct platform interventions). The emergence of the `\textit{Debunk-It-Yourself}' approach thus might be seen as a threat both to the profitability of the platforms and national security interests of the governments, leaving misinformation unaddressed in the long run. While we see this scenario as realistic and probable, we nonetheless believe that the `\textit{Debunk-It-Yourself}' approach has a potential to finds its way quickly to health misinformation, hopefully across platforms in near future. Doing so, we envision that the `\textit{Debunk-It-Yourself}' will evolve to a state where nuanced dissemination of manipulative narratives with intention (i.e., disinformation and any broader information operations campaigns) is early on detected, debunked, and collaboratively (hopefully) contained. Given that past disinformation operations, disinformation and propaganda function as collaborative ``work'' within online crowds on social media platforms~\cite{Starbird2019}, we see the `\textit{Debunk-It-Yourself}' naturally pairing this process to counter nefarious mass influence efforts.

% benefits that allow for new platforms, able to struck a profitable yet trustworthy content dissemination, to emerge. Obviously, the case of TikTok and the emergence of `\textit{Debunk-It-Yourself}' due to the unique platform features of ``stich'' and ``duet'' 

% Individual debunkers, as the participants in our study, might be equally beneficial as ``boots on the ground" able to independently identify potential disinformation and information operation campaigns proactively, in contrast to the reactive effort post 2015 UK's Brexit referendum and 2016 US Presidential elections.  

% Evoking the case of privacy invasion and mass surveillance exposed by Edward Snowden, the question arises whether and to what extend the government is prepared to trade their own citizens' ability to obtain true and trustworthy information for advantage in the network-of-intelligence game, allowed through access to social media platforms, including TikTok \cite{anderson2014privacy}.

The sustainability of the `\textit{Debunk-It-Yourself}' model is also influenced by the dynamic nature of social media platforms like TikTok. Factors such as changes in monetary incentives or platform policies, and the motivation to confront falsehoods play significant roles in either promoting or curtailing the spread of misleading claims. We didn't study the opportunity costs involved for the debunkers nor the costs involved in creating and maintaining a ``debunker persona'' on TikTok. These costs include establishing and maintaining parasocial interactions with the `\textit{Debunk-It-Yourself}' audience, maintaining persuasive tone and approach in creating the videos, and maintaining a distance from the actual ``influencers'' that promote brands for personal benefit~\cite{Willis_Delbaere_2022}. The ``promotion'' of science or scientific facts is a costly endeavor as demands keeping abreast with the latest research and advancement in rapid fields such as mental health and fitness/nutrition, where the knowledge is far from settled. As mental health and fitness/nutrition are the most pronounced topics that shape the social inclusion on TikTok, the `\textit{Debunk-It-Yourself}' practitioners need also to assume the cost of ``influencing'' the ``influencers' to maintain relevance and credibility especially across younger TikTok users~\cite{Motta2024}.

Additionally, political or regulatory shifts, such as the proposed US ban on TikTok, could lead to abrupt changes in the platform's operations, potentially necessitating a swift migration of content creators to alternative platforms~\cite{Maheshwari2024}. The resilience and adaptability of `\textit{Debunk-It-Yourself}' practitioners to these changes will be pivotal in maintaining the initiative's effectiveness. Here, it is important to pay equal attention to the attrition among individual debunkers on the platform. The monetary incentives and platform policies could do so much to aid the individuals' willingness to keep debunking, if, for example, they give up in face of increased misinformation of direct harassment from other users on the platform. 

While the individual debunkers could report such a behavior, currently there are no statutory interventions on TikTok, nor general regulations in the US (where our study was based) that particularly protect individuals who render such an important service of truth arbitration to the wider community. The willingness of the platform to protect individual debunkers will signal, at least initially, that the `\textit{Debunk-It-Yourself}' has an intrinsic social responsibility value, and with that might incentivize broader participation. As such, the `\textit{Debunk-It-Yourself}' initiative makes a good case for exploring regulatory interventions to incentivize TikTok (and other platforms) to do more proactive information debunking and prevent a race-to-the-bottom effect where an affective and entertaining content drowns down facts and trustworthy health practices of equal importance for individual as well as for public health. This social value, taken together with the proliferation of health misinformation across platforms, suggests that the principles of `\textit{Debunk-It-Yourself}' could be beneficial if adapted for use on platforms other than TikTok. The development of platform-agnostic debunking strategies and statutory protections for individual debunkers  could ensure the continuity and effectiveness of misinformation countermeasures across the digital landscape.

\subsection{Limitations and Future Work}
Our research focused on the debunking of health misinformation on TikTok by professionals in the mental health and fitness/nutrition fields. We must note here that some of the videos on these two topics targeted by the debunkers might be subjectively considered as ``misleading'' or ``manipulative'' as there are gray areas and subject of emergent evidence and scientific facts. Here, we don't exclude the possibility that two  `\textit{Debunk-It-Yourself}' practitioners might have conflicting views on a particular medical issue and diverge in their choice to debunk not. Although we structured our inquiry to follow the ``Debunking Handbook,'' there might exist other approaches for countering misleading medical advice and we hope to revisit the `\textit{Debunk-It-Yourself}' according to these in future. Here, we acknowledge all the ``debunking'' alternatives depend on the virality as a critical engagement factor, be that for viewing-driven (TikTok, Instagram) or interaction-driven participation (Facebook, Twitter, etc..), something that we took it as an assumption but it certainly begets a study of its own. 

Although our sample was balanced and representative within our research scope, exploring broader or differently composed samples could yield different conclusions, a direction we plan to pursue in future studies. We were limited by the extensive recruitment vetting process as the meticulous and thorough approach to check the credentials and/or accreditation of each individual debunkers, albeit very rigorous, it nonetheless precluded our feasibility to reach out to a bigger population of individual debunkers. Additionally, we limited our sampling to English-language videos, acknowledging that different languages and regional content might influence the interaction with social media recommendation algorithms and engagement methods, potentially affecting the applicability of our findings across various linguistic and cultural contexts~\cite{Boeker2022}. 

Furthermore, our analysis did not assess the actual impact of the debunking videos, an area we aim to explore in subsequent research. Our study is limited by assuming a default adversarial model of misinformation without assigning a specific actor behind, given that the misleading videos that we sampled in our dataset have been mostly driven by the creator rewards program by TikTok~\cite{tiktok-creator-rewards}. Equally, we haven't delved deep in to the nuanced of the defender's model employed by TikTok, assuming the position of minimal involvement through platform-based moderation. We haven't looked for indicators of intentional misinformation dissemination, nor we looked into any occurrences where a debunked video is stitched again by the original creator to respond to the debunking. We beleive that these are studies of their own and we do plan to pursue each of them in the near future.

We also recognize the limitations imposed by the current state of TikTok's moderation policies, which are subject to change and could impact future debunking practices. While we assumed a minimal involvement, we cannot say that that would be the approach TikTok takes in future and --- including changes in participation affordances --- decides to eliminate any misinformation \textit{en masse}, effectively rendering the `\textit{Debunk-It-Yourself}' effort of no use. Our decision to exclude politically related content means that our results may not fully reflect the complexities of misinformation, which often intertwines with political issues. We not yet considered the audience angle of the debunking process or the reception and the dialogue between the TikTok user base and the debunkers, especially among the younger part of the population. Their reception, response, dialogue, and actions stemming from the interaction with the `\textit{Debunk-It-Yourself}' practitioners (and both the debunking and original content) undoubtedly shapes the present and future participation and we plan to continue our study of the debunking process using sociotechnical and collaborative lenses.

In addition to the assessment of the effect of the debunking videos with the users, we plan to broaden our inquiry to include controlled politically related content, especially in the run up period during the US 2024 Elections in order to gain further insights into the the robustness of the `\textit{Debunk-It-Yourself}' model across topics. Our agenda also includes a longitudinal or diary study where we would focus on the feedback the individual debunkers get from each of their videos and how this feedback informs future debunking efforts is also on our agenda. Drawing on the findings we presented in Section 5.8, the goal of such a study would be to illuminate how the `\textit{Debunk-It-Yourself}' approach evolves around one particular topic, and how the knowledge of this evolution (and the accompanying misinformation) could inform any potential change in the TikTok's moderation strategies as well as more aggressive government regulatory effort. For example, an open research question is whether TikTok penalizes these individual debunkers every time they report a misinformation video (e.g., ``shadowban''), investigates the video containing the claims further (and takes actions to promote/demote it),  Another open research question is whether there is a shift in strategy based on the perceived platform treatment or any feedback given in the comments section of the debunking video (positive or negative). Our study takes a cumulative view up to the point of the data collection but that is insufficient to capture any potential shift in debunking strategies that may occur in light of new misinformation, changes in TikTok individual user treatment, moderation strategies, or viewer' feedback, all of which are important  to be studied in order to better understand the full potential of the `\textit{Debunk-It-Yourself}' approach.

\section{Conclusion}
% The ``sticky'' of misinformation on TiTok, reinforced by the algorithmic curation of videos containing questionable content relative to nutrition or mental health, was a sufficiently potent challenge for health professionals to devise a response in kind. Undoing the stickiness was done through a `\textit{Debunk-It-Yourself}' process where videos explaining the falsity of questionable content and offering scientifically facts were directly stitched or tagged to ones deemed harmful to individual health. A step away from the traditional approach of moderation, the `\textit{Debunk-It-Yourself}' offers a platform-independent avenue for misinformation counter-influence that -- instead of relying on warnings or third party content -- does it through actual videos on TikTok. While facing the usual challenges of subjectivity in selection, scalability, or going against the preferred, viral nature of social media content, it does appear promising in the effort to drown the truth with what appears to be truth not just on TikTok but all mainstream platforms. And not just about misinformation relative to individual health, but also relative to public health, societal order, synthetic media, and perhaps even electoral affairs. 
The pervasive \lq\lq stickiness\rq\rq~of misinformation on TikTok, amplified by algorithmic promotions of videos with dubious claims about nutrition or mental health, presents a significant challenge for health professionals and poses potential consequential impacts on individuals as well as the society as a whole. In response, these healthcare professionals developed a ``\textit{Debunk-It-Yourself}'' strategy, where videos debunking false claims and presenting scientific facts are directly linked or stitched to those spreading harmful misinformation. Our study, involving $14$ health professionals, is the first to examine this topic, demonstrating how such approaches by the \textit{debunkers} mark a departure from traditional moderation techniques and provide a platform-independent response mechanism for countering misinformation through direct engagement on TikTok. Despite facing challenges such as subjectivity in video selection, scalability, and resistance due to the inherently viral nature of social media content, this strategy shows promise in addressing not only health-related misinformation but also broader issues affecting public health, societal stability, synthetic media, and potentially electoral processes across mainstream platforms. 
%Our work highlights these findings, creating opportunities for future research and growth in this area.

\bibliographystyle{ACM-Reference-Format}
\bibliography{TikTokMisinformationDebunkers/main}

% \newpage
\appendix
\newpage
\appendix

\section{Study Questionnaire} \label{app:study-questions}
\begin{enumerate}

\item What social media platforms do you use to debunk videos?

\item How long have you been doing the debunking videos on each platform?

\item About how many debunking videos have you created?

\item What topic(s) are you debunking videos on?

\item How do you find videos to debunk (please explain to us your criteria respective to whether you respond to blatant falsehoods, rumors, etc)?

\item How do you decide to which video(s) to respond to and what content to create in response?	

\item Do you base your debunking based solely on your experience or do you also consult fact-checking services, academic journals, or other resources?	

\item How do define misinformation in your own words?

\item How familiar are you with the algorithms used by social media platforms - both for recommendation and for removing/labeling misinformation? Please expand on what your knowledge of the subject.

\item How familiar are you with the strategies used by influencers to expand their number of followers and views (and their awareness of the recommendation and misinformation handling algorithms)? Please expand on your knowledge of the subject.

\item What do you feel makes the misinformation videos successful in disseminating inaccurate information?

\item What tactics do you use to have people view your video instead of the original that contains misinformation (tag it, duet, comments, others)?

\item If a video has a warning label, does this impact your decision to create a duet with it?	

\item Do you also report a video that you debunk to the platform administrators as a ``misinformation''?

\item Have the social media platforms moderated any of your videos using soft moderation, such as warning labels, or removed them? If yes, what was your experience with this and what action did you take?

\item Have you identified any other content creators that you view as a role model for your videos or do you collaborate with other debunkers like yourself? If yes, what qualities do you feel make their videos effective?	

\item Do you think that the platform could leverage your work and if so in what capacity towards curbing the spread of misinformation?	
\end{enumerate}

\section{Codebook} \label{app:codebook}

% MISINFORMATION INITIATION
% Q5 - How do you decide to which video(s) to respond to and what content to create in response?

    \begin{itemize}
        \item \textbf{Misinformation Debunking Initiation} -- Codes pertaining to the initiation approach for debunking misinformation videos
        
        \begin{itemize}
            \item \textbf{Recommended by TikTok} -- The participant expressed they initiate their DIY debunking for videos recommended to them by TikTok on their ``For You'' page;  
            
            \item \textbf{Perceived danger of a misinformation claim} -- The participant expressed they initiated their DIY debunking based on what they perceived as the most dangerous misinformation that needed to be immediately debunked;  
            
            \item \textbf{Tagged by followers} -- The participant expressed they initiated their DIY debunking when they were tagged by their followers in a video that made a misinformation claim;
            
            \item \textbf{Based on requests} -- The participant expressed they initiated their DIY debunking based on requests or inquiries from students or patients.

        \end{itemize}
    \end{itemize}

% MISINFORMATION SELECTION
% Q6 - How do you find videos to debunk (please explain to us your criteria respective to whether you respond to blatant falsehoods, rumors, etc)?

    \begin{itemize}
        \item \textbf{Misinformation Debunking Selection} -- Codes pertaining to the approach for selecting misinformation videos for direct DIY debunking
        
        \begin{itemize}
            \item \textbf{Virality} -- The participant expressed they select the videos they want to debunk based on the videos' number of views and followers (the content creator of the video has);
            
            \item \textbf{Susceptibility -- Comments' Agreement} The participant expressed they select the videos they want to debunk based on perceived agreement with the claim in the video expressed in the comments by other TikTok users (susceptibility to misinformation);
            
            \item \textbf{Blatant Falsehoods} -- The participant expressed they select the videos they want to based on what they perceived as the video containing the most blatant falsehoods at the moment that needed to be immediately debunked.

        \end{itemize}
    \end{itemize}

% MISINFORMATION DEBUNKING
% Q7 - Do you base your debunking based solely on your experience or do you also consult fact-checking services, academic journals, or other resources?
    
    \begin{itemize}
        \item \textbf{Misinformation Debunking} -- Codes pertaining to the direct DIY debunking approach taken by the participants
        
        \begin{itemize}
            \item \textbf{Scientific evidence} -- The participant expressed they rely on scientific evidence to debunk misinformation;
            
            \item \textbf{Clinical experience} -- The participant expressed they rely on their clinical experience to debunk misinformation;

            \item \textbf{Consultation with peers} -- The participant expressed they rely on consultation with peers to debunk misinformation;
            
        \end{itemize}
    \end{itemize}

% MISINFORMATION DEFINITION
% Q8 - How do define misinformation in your own words?
    
    \begin{itemize}
        \item \textbf{Misinformation Definition} -- Codes pertaining to the definition of what constitutes as ``misinformation'' on TikTok in the view of our participants
        
        \begin{itemize}
            \item \textbf{Falsehoods with \textit{no} intent to deceive -- misinformation} -- The participant defined misinformation as videos on TikTok containing false claims, spread without direct intent to deceive the viewers or alter the information space on a particular topic;
            
            \item \textbf{Falsehoods with intent to deceive -- disinformation} -- The participant defined misinformation as videos on TikTok containing false claims, spread with a direct intent to deceive the viewers and alter the information space on a particular topic;

        \end{itemize}
    \end{itemize}

% ALGORITHMIC FAMILIARITY
% Q9 - How familiar are you with the algorithms used by social media platforms - both for recommendation and for removing/labeling misinformation? Please expand on what your knowledge of the subject.
    
    \begin{itemize}
        \item \textbf{TikTok Algorithmic Familiarity} -- Codes pertaining to the familiarity with the workings of the TikTok's recommendation and moderation algorithms 
        
        \begin{itemize}
            \item \textbf{Unfamiliar} -- The participant expressed they are unfamiliar how the TikTok's recommendation and moderation algorithms works;
            
            \item \textbf{Favoring sensationalism and divisiveness} -- The participant expressed they see the TikTok's recommendation algorithm favoring sensationalist and divisive content;

            \item \textbf{Favoring engagement} -- The participant expressed they see the TikTok's recommendation algorithm promoting videos based off of engagement;

            \item \textbf{Lax Moderation} -- The participant expressed they see the TikTok's moderation algorithm rarely removes videos with misinformation claims;  

            \item \textbf{Unfairness} -- The participant expressed they see the TikTok's moderation algorithm removing videos that debunk misinformation claim and on the expense of  promoting videos with misinformation claims by  the TikTok's recommendation algorithm; 
            
        \end{itemize}
    \end{itemize}

% INFLUENCER STRATEGIES
% Q10 - How familiar are you with the strategies used by influencers to expand their number of followers and views (and their awareness of the recommendation and misinformation handling algorithms)? Please expand on your knowledge of the subject.

% ATTENTION CHECK - Q11 - What do you feel makes the misinformation videos successful in disseminating inaccurate information?

    \begin{itemize}
        \item \textbf{Influencer Strategies} -- Codes pertaining to the familiarity with the influencer's strategies for increasing engagement (number of followers and viewers) 
        
        \begin{itemize}
            \item \textbf{Fear-mongering} -- The participant expressed they see influencers use fear-mongering as a strategy to gain followers and views for their content;
            
            \item \textbf{Controversy} -- The participant expressed they see influencers use controversial claims and narratives as a strategy to gain followers and views for their content;

            \item \textbf{Shock / emotion appeal} -- The participant expressed they see influencers use shocking or emotionally appealing claims and narratives as a strategy to gain followers and views for their content;
                        
        \end{itemize}
    \end{itemize}

% OWN INFLUENCINT TACTICS
% Q12 - What tactics do you use to have people view your video instead of the original that contains misinformation (tag it, duet, comments, others)?

    \begin{itemize}
        \item \textbf{Own Influencing Tactics} -- Codes pertaining to the participants' own influencing tactics for increasing engagement (number of followers and viewers) 
        
        \begin{itemize}
             \item \textbf{Duets/Stitch} -- The participant expressed they use the duet or stitch feature together with the video containing the claim they are debunking as a tactic to gain followers and views for their content;

             \item \textbf{Tagging} -- The participant expressed they use the tag the video containing the claim they are debunking as a tactic to gain followers and views for their content;

             \item \textbf{Hashtags} -- The participant expressed they use the similar hashtags the video containing the claim they are debunking contains as a tactic to gain followers and views for their content;

            \item \textbf{Frequently posting} -- The participant expressed they frequently post their own content as a tactic to gain followers and views for their content;

        \end{itemize}
    \end{itemize}

% WARNING LABEL
% Q13 - If a video has warning label, does this impact your decision to create a duet with it?

    \begin{itemize}
        \item \textbf{Warning Label on Videos} -- Codes pertaining to the participants' own debunking tactics in presence of a warning label attached to videos by TikTok
        
        \begin{itemize}
             \item \textbf{Proceed} -- The participant expressed they proceed with their debunking even if TikTok has attached a warning label to a video contain the claim they are responding to;

             \item \textbf{Abstain} -- The participant expressed they abstain from debunking videos to which TikTok has attached a warning label;

             \item \textbf{Claim dependent} -- The participant expressed they proceed with their debunking of videos that TikTok has attached a warning label to them if the claim is convincing and the response in the comments indicate other users are susceptible to the misleading claim;

            \item \textbf{Hasn't seen warning labels} -- The participant expressed they have not seen warning labels on TikTok or haven't seen them on the type of content they debunk. This is not something they have encountered and do not anticipate encountering it.

        \end{itemize}
    \end{itemize}

% REPORTING
% Q14 - Do you also report a video that you debunk to the platform administrators as a ‘’misinformation’’?

    \begin{itemize}
        \item \textbf{Reporting} -- Codes pertaining to the participants' own debunking tactics involving reporting of videos containing misinformative claims
        
        \begin{itemize}
             \item \textbf{Yes} -- The participant expressed they report TikTok videos they believe contain a misinformation claim;

             \item \textbf{No} -- The participant expressed they don't report TikTok videos they believe contain a misinformation claim;

             \item \textbf{Claim dependent} -- The participant expressed they report TikTok videos they believe contain a misinformation claim only if the claim is convincing and the response in the comments indicate other users are susceptible to the misleading claim;

        \end{itemize}
    \end{itemize}

% MODERATION
% Q15 - Have the social media platforms moderated any of your videos using warning labels, or removed them? If yes, what was your experience with this and what action did you take?

    \begin{itemize}
        \item \textbf{Being Moderated} -- Codes pertaining to the participants' experience relative to their content being assigned a warning label by TikTok (soft moderation) or removed (hard moderation)
        
        \begin{itemize}
             \item \textbf{Wrongly Moderated} -- The participant expressed their content was moderated by TikTok but the warning label was removed after they complained about it;

             \item \textbf{Shadow-banned} -- The participant expressed that TikTok shadow-banned content (only pushed it out to followers instead of through hashtags, audio, or other places where it could organically show up for non-followers);

             \item \textbf{No} -- The participant expressed their content has never been subject of moderation.

        \end{itemize}
    \end{itemize}

% DIY DEBUNKING ROLE MODELS/COLLABORATION
% Q16 - Have you identified any other content creators that you view as a role model for your videos or do you collaborate with other debunkers like yourself? If yes, what qualities do you feel make their videos effective?

    \begin{itemize}
        \item \textbf{DIY Debunking Collaboration} -- Codes pertaining to the participants' perspectives relative to role models and collaboration with other content creators who debunk videos they believe contain misinformation claims on TikTok
        
        \begin{itemize}
             \item \textbf{Indirect collaboration} -- The participant expressed they are looking for role model creators (attributes:  succinct and smart rebuttal in a very short amount of time; delivers information compassionately; leaves room for the nuanced conversation; supports their information with sources when necessary);

             \item \textbf{Direct collaboration} -- The participant expressed that they are open for collaboration with other content creators that individually debunk misinformation on TikTok; 

             \item \textbf{No collaboration} -- The participant expressed that they are not open for collaboration with other content creators that individually debunk misinformation on TikTok.

        \end{itemize}
    \end{itemize}

% TIKTOK LEVERAGING DIY DEBUNKING 
% Q17 - Do you think that the platform could leverage your work and if so in what capacity towards curbing the spread of misinformation?

   \begin{itemize}
        \item \textbf{Leveraging DIY Debunking} -- Codes pertaining to the participants' perspectives relative to the opportunity for TikTok to leverage their DIY debunking efforts for curbing misinformation
        
        \begin{itemize}
             \item \textbf{Prioritize their content} -- The participant expressed that they believe TikTok could prioritize their content on controversial topics; 

             \item \textbf{Compensate them} -- The participant expressed that they believe TikTok could compensate them for their work;

             \item \textbf{No leveraging} -- The participant expressed that they don't think TikTok could leverage their work.

        \end{itemize}
    \end{itemize}

\end{document}